\documentclass[aps,pra,showpacs,reprint,superscriptaddress,]{revtex4-2}
\usepackage{graphicx}% Include figure files
\usepackage{dcolumn}% Align table columns on decimal point
\usepackage{bm}% bold math
\usepackage{amsmath, amssymb, mathtools, isomath, nccmath}
\usepackage{siunitx}
\usepackage{xspace}
\usepackage{xcolor}
\usepackage{siunitx}
\usepackage[british]{babel}
\usepackage{sidecap}

\usepackage[colorlinks,
            linkcolor=red,     
            anchorcolor=blue,  
            citecolor=blue,      
            ]{hyperref}

\usepackage{xcolor}

\begin{document}

\title{Entanglement transition through Hilbert-space localization}

  \author{Quancheng Liu} 
  \affiliation{Department of Physics, Institute of Nanotechnology and Advanced Materials, Bar-Ilan University, Ramat-Gan 52900, Israel}
   \author{Klaus Ziegler}
    \thanks{klaus.ziegler@physik.uni-augsburg.de}
   \affiliation{Institut f\"ur Physik, Universit\"at Augsburg, D-86135 Augsburg, Germany}
   
\date{\today}

\begin{abstract}
We study Hilbert-space localization of the many-body dynamics due to ergodicity breaking and analyze this effect in terms
of the entanglement entropy and the entanglement spectrum. We find a transition from a regime driven by quantum 
tunneling to a regime that is dominated by boson-boson interaction, where the latter exhibits ergodicity breaking.
Properties of this transition are captured by observation time averaging, which effectively suppresses the large dynamical entanglement fluctuations near the critical point. We employ this approach to the experimentally available Bosonic 
Josephson Junction. 
In this example the transition from a tunneling regime to Hilbert-space localization reveals clear signatures in the 
entanglement entropy and entanglement spectrum. Interestingly, the transition point is reduced by quantum effects in comparison to the well-known result of the mean-field approximation in the form of self-trapping. This indicates that
quantum fluctuations reduce the classical self-trapping. 
Different scaling with the respect to the number of bosons $N$ are found in the tunneling and the localization regime: 
While the entanglement entropy grows 
{\it logarithmically} with $N$ in the tunneling regime, it increases {\it linearly} in the localized regime. 
Our results indicate that entanglement provides a concept for a sensitive diagnosis for the transition
from a quantum tunneling regime to Hilbert-space localization.
\end{abstract}

\maketitle

%{\it Introduction.---}

\section{Introduction}

Entanglement is considered as one of the most fundamental building blocks in quantum physics and quantum information processing. Specifically, the R\'enyi entanglement entropy measures the quantum correlations between two subsystems under a spatial bipartition~\cite{PhysRevD.34.373,PhysRevLett.71.666,RevModPhys.82.277,PhysRevLett.127.040603}, which has become an important and popular concept for detecting measurement-induced entanglement transitions~\cite{PhysRevB.100.134306,PhysRevX.9.031009,PhysRevB.104.155111}, characterizing many-body dynamics and localization~\cite{amicoRevModPhys.80.517,RevModPhys.91.021001,kitaevPhysRevLett.90.227902,calabrese2004,peschel2009,calabrese2009,PhysRevX.5.041047,PhysRevX.8.021062,PhysRevX.8.041019,doi:10.1146/annurev-conmatphys-031214-014726}, and classifying the topology of quantum systems~\cite{PhysRevLett.96.110405,PhysRevLett.105.080501,Jiang2012,Miao2022eigenstate}. Recently, the R\'enyi entanglement entropy has been efficiently measured in the laboratory with randomized methods without full quantum state tomography~\cite{PhysRevLett.108.110503,PhysRevLett.120.050406,doi:10.1126/science.abi8378}.

Hilbert-space localization (HSL) describes the phenomenon in which the evolution of a quantum system from an initial Fock
state is restricted to a subregion of the Hilbert space in the presence of strong particle-particle interaction; 
i.e., it is a special form of spontaneous ergodicity breaking. It originates from the fact that a Fock state is an eigenstate
of the local interaction part of the Hamiltonian, where only a sufficiently strong tunneling can overcome the HSL.  
Although we anticipate such a behavior for any system in which tunneling competes with
particle-particle interaction, so far it was only calculated for the Bosonic Josephson Junction (BJJ).
In this case we have seen that it is characterized by a change of the scaling behavior of the participation
ratio~\cite{cohen16} or by a sudden jump of the return probability between different states \cite{symmetry21}.
The dynamics of the system is rather complex, where the scattering of an individual particle by other particles can 
be considered as random. Therefore, HSL is reminiscent of Anderson localization in real space~\cite{PhysRev.109.1492},
with the crucial difference that the scattering environment in the latter is static rather than dynamic.
It also distinguishes itself from other types of many-body localization, where the interplay of interactions and disorder is involved~\cite{amicoRevModPhys.80.517,PhysRevB.88.014206,RevModPhys.91.021001}. In this Letter, we focus 
on the BJJ, which has been studied intensively in the semiclassical 
limit~\cite{PhysRevA.55.4318,PhysRevA.82.053617,PhysRevLett.125.134101,PhysRevA.103.023326}, and 
realized experimentally~\cite{PhysRevA.80.053613,Abbarchi2013,PhysRevLett.118.230403,PhysRevLett.120.173601,PhysRevResearch.3.023197}. It is found that the entanglement transition and different properties 
are strongly related to the HSL.

The implementation of entanglement in the investigation of many-body localization has led to some of the most important discoveries in many-body physics. For example, the scaling laws of R\'enyi entanglement entropy are taken as the key characteristic of the many-body localization phase. Nevertheless, it is still unclear how the entanglement behaves when a many-body system undergoes the HSL phase transition, arising from the competition of particle interactions and tunneling. Here we investigate the following questions: Is there a generic entanglement transition and HSL correspondence? Will the entanglement exhibit different characteristic properties in the localized and tunneling phases? Can entanglement reveal quantum effects that are not covered by the classical mean-field approximation? Affirmative answers are obtained in this work. 

The unitary evolution of a quantum system typically drives it to states of higher entanglement~\cite{PhysRevLett.111.127205,PhysRevX.9.031009} and the dynamical entanglement entropy usually exhibits large fluctuations, which makes it difficult to extract generic information about the system. Here we adopt the principle of ensemble average from statistical physics for the ensemble created by randomly chosen observation times. We denote this approach as the observation time average method. In the experiment, this means one repeats the measurement at different times on identical systems and averages over this ensemble of results. This idea has some similarities with the Random Matrix Theory, where averaging an ensemble of Hamiltonians is used to obtain generic properties of many-body spectrum~\cite{10.2307/1970079,PhysRev.104.483,doi:10.1063/1.1703773,10.2307/2027409,mehta2004random,RevModPhys.69.731}. It is shown that the observation time average method is a powerful tool and efficiently suppresses the randomness in entanglement entropy and entanglement spectrum. Moreover, this approach can be extended to the investigation of other systems with entanglement.

\begin{figure}
    \centering
    \includegraphics[width=\linewidth]{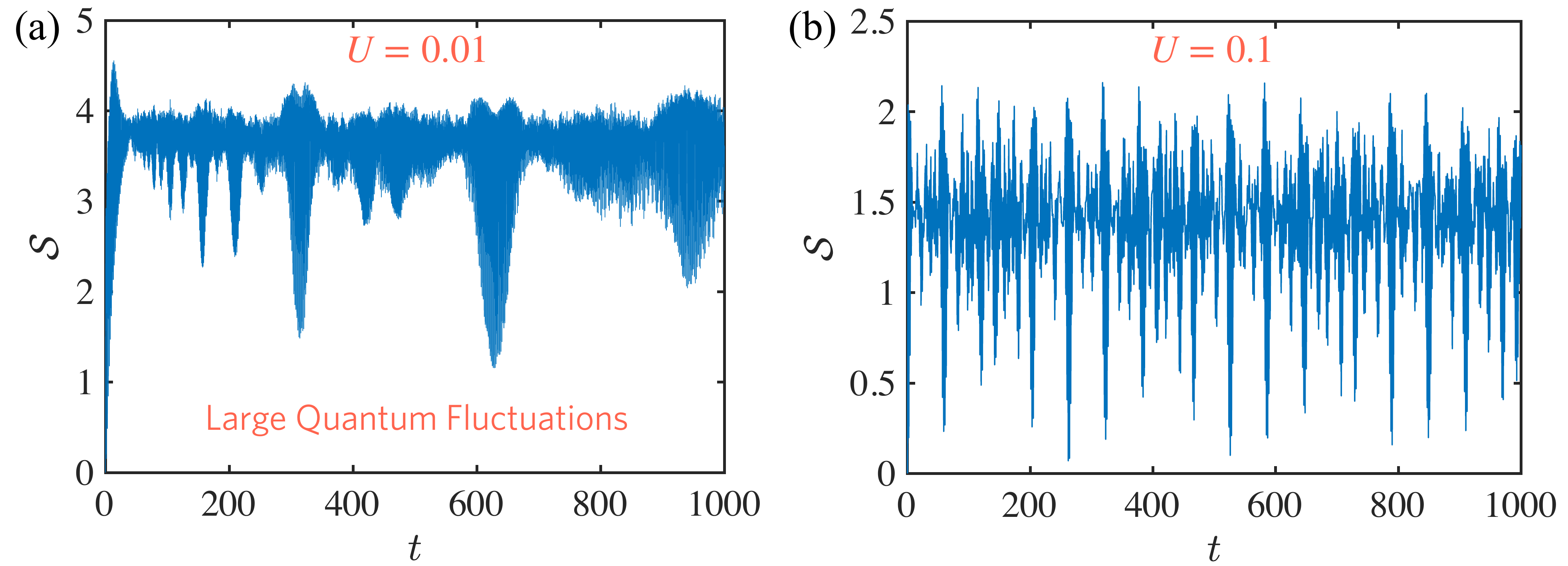} 
    \caption{The entanglement entropy ${\cal S}$ for 100 bosons as a function of time $t$ for  $U=0.01$ (a) and $U=0.1$ (b). We set $J=1$. As shown in the figure, the entanglement entropy fluctuates strongly in time, which is later suppressed with the observation time average.   
    }    
    \label{fig:1}
\end{figure}

%{\it Model.---}
\section{Model}

Within the single mode approximation~\cite{PhysRevA.55.4318}, 
the BJJ with $N$ bosons can be described as a two-site Bose-Hubbard model,
\begin{equation}
	H_{{\rm BJJ}}=-J(a_L^\dagger a_R + a_R^\dagger a_L)+ \frac{U}{2}(n_L^2+n_R^2),
\label{ham00}
\end{equation}
where $a^\dagger_{L,R}$ ($a_{L,R}$) are the bosonic creation (annihilation) operators in the left/right potential traps, and $n_{L,R}=a_{L,R}^\dagger a_{L,R}$ are the corresponding number operators. $J$ describes the tunneling of bosons between the traps, and $U$ represents the particle-particle interaction, which favors energetically a symmetric distribution of bosons in the double traps when $U>0$. Using Fock states $|k,N-k\rangle \equiv |k\rangle\otimes|N-k\rangle$ ($k=0,\cdots,N$) as a basis of the Hilbert space, the corresponding Hamiltonian matrix has a tridiagonal structure with $ H_{k,k'}=\langle k,N-k|H_{{\rm BJJ}}|k',N-k'\rangle= U[(N-k)^2+k^2] \delta_{k,k'}  -J\sqrt{k(N+1-k)} \delta_{k,k'-1}  -J\sqrt{k'(N+1-k')} \delta_{k,k'+1} $. This matrix can be interpreted as a $(N+1)$-site tight-binding lattice with broken translational invariance, where the tunneling rate and the potential are minimal at the center and grow symmetrically towards the endpoints. Hence, the HSL in the BJJ is related to translation symmetry breaking, in contrast to Anderson localization in the presence of quenched disorder~\cite{PhysRev.109.1492}, whose ensemble is translational invariant. Using the $SU(2)$ spin representation, the BJJ Hamiltonian can also be written as~\cite{PhysRevA.55.4318}
\begin{equation}
	H_{{\rm BJJ}}^{S} \ = \ UL_z^2 - 2JL_x +UN^2/4,
\label{spin_ham}
\end{equation}
when $L_x = (a_L^\dagger a_R + a_R^\dagger a_L)/2$ and $L_z = (a_L^\dagger a_L-a_R^\dagger a_R)/2$, representing a large nonlinear spin system with magnitude $S=N/2$. This maps the evolution of BJJ to the spin motion on the Bloch sphere.

\begin{figure*}
    \centering
    \includegraphics[width=1\linewidth]{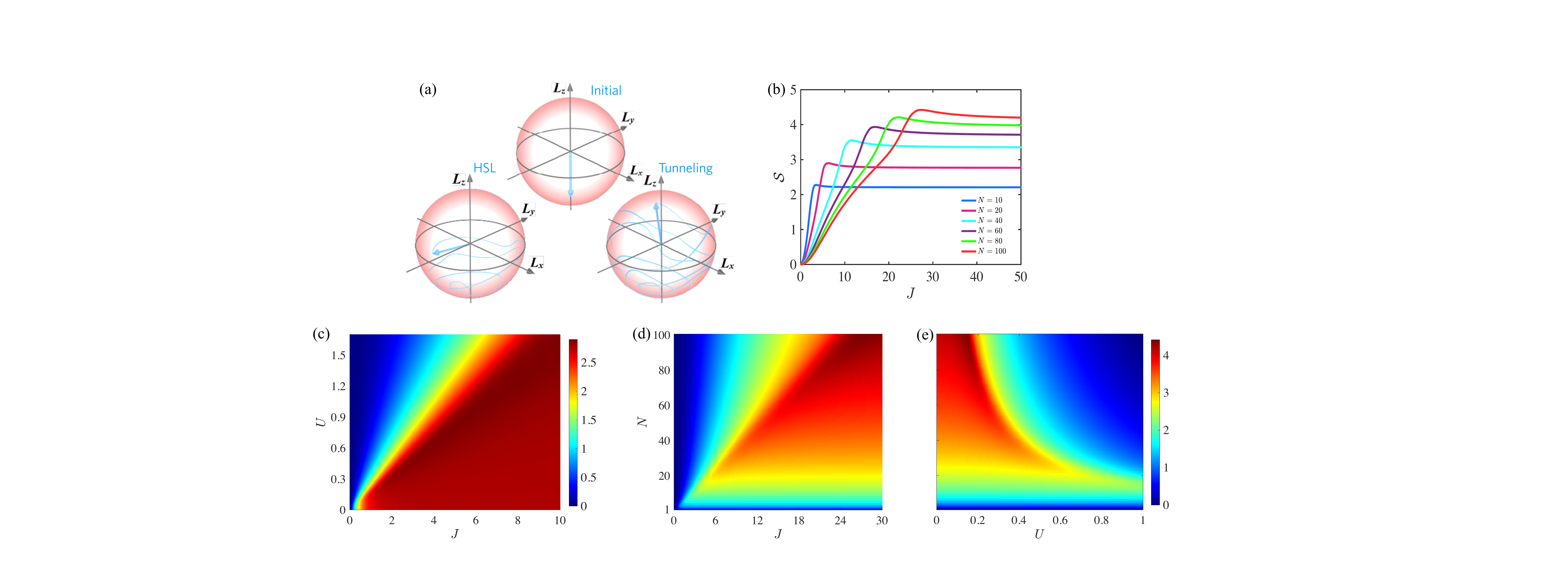}
    \caption{The entanglement entropy phase transition with system parameters. (a) Schematic plot of the HSL phase and tunneling phase on the Bloch sphere. In (b), we plot ${\cal S}(J)$ for different $N$s with $U=1$. When $J$ is small, corresponding to the HSL phase, ${\cal S}$ increases with $J$. The system undergoes the phase transition to the tunneling regime at $J_c$, after which ${\cal S}$ is a constant. The competition of $U$ and $J$ is shown in (c), where we choose $N=20$ and the colors represent the magnitude of ${\cal S}$. In (d,e), we present the effects of the number of bosons on the phase transition. Specifically,  we plot ${\cal S}(J, N)$ with $U=1$ in (d) and ${\cal S}(U, N)$ with $J=1$ in (e). All four plots (b)-(e) indicate a clear entanglement phase transition when the system undergoes the transition from the HSL phase to the tunneling phase controlled by the $U, J$, and $N$.}    
    \label{fig:2}
\end{figure*}

\begin{figure}
    \centering
    \includegraphics[width=\linewidth]{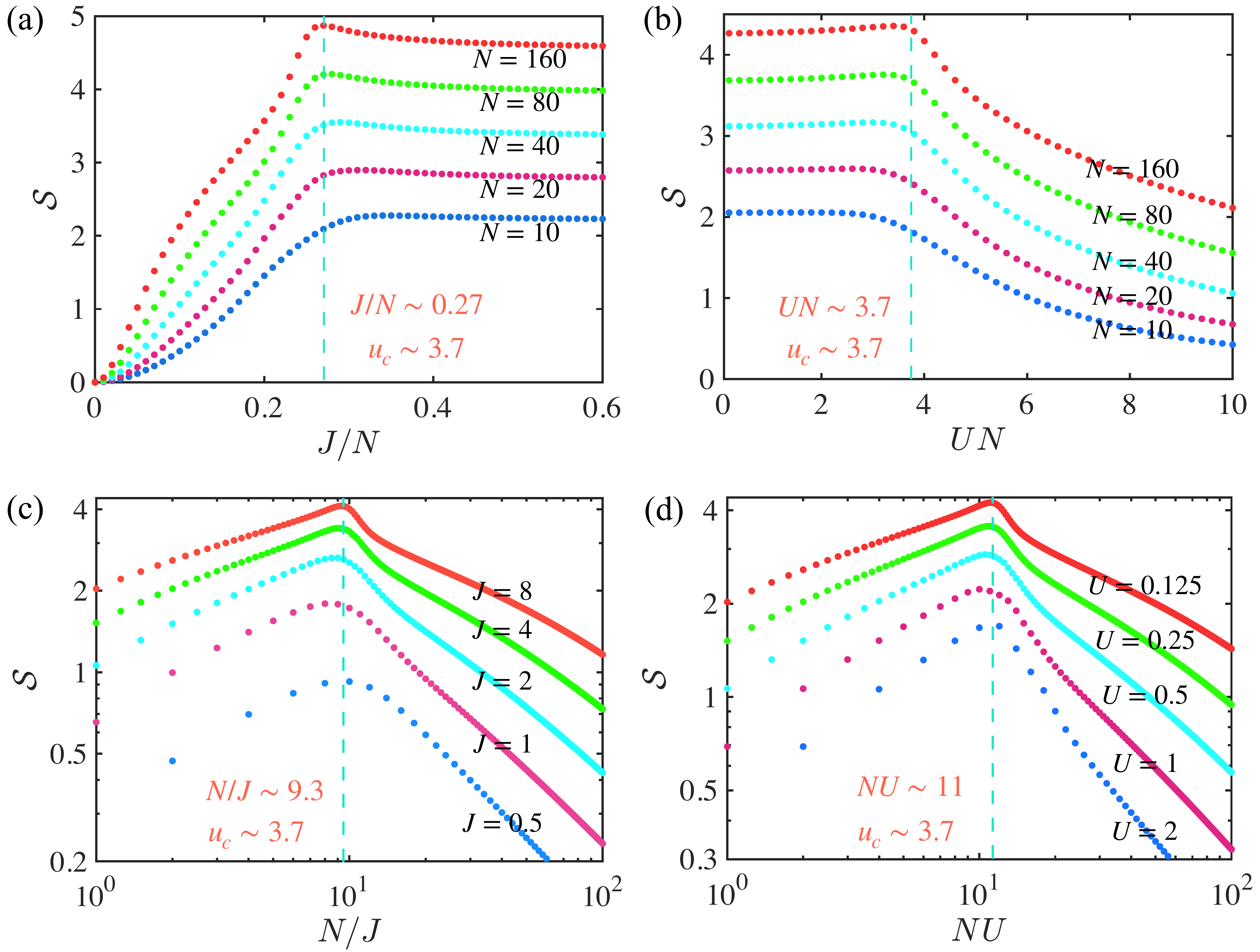}
    \caption{The characterization of the critical transition point $u_c$. In (a), we plot ${\cal S}$ versus $J$ rescaled by $N$ with $U=1$. Similarly, we plot ${\cal S}(UN)$ with $J=1$ in (b). The phase transition point $u_c \sim 3.7$ for all the realizations. We plot the dependence of the entanglement entropy with $N$ rescaled by $J^{-1}$ with $U=0.4$ in (c) and rescaled by $U$ with $J=3$ in (d). Again, the entanglement entropy undergoes a phase transition at $u_c \sim 3.7$ and reaches the maximum at this critical value. Hence, $u_c$ is a generic value that characterizes the entanglement phase transition.}    
    \label{fig:3}
\end{figure}

\section{Entanglement and R\'enyi entropy}

The Hilbert space ${\cal H}$ of the BJJ is a product space ${\cal H}_l\otimes{\cal H}_r$ that comprises the left and right traps. We start with all bosons in the right well, i.e., $\rho_0 = |0,N\rangle\langle0,N|$. On the Bloch sphere, the initial spin vector $\langle 0,N|{\bf L}|0,N\rangle$ (${\bf L}=(L_x,L_y,L_z)$) points to the south pole [Fig. \ref{fig:2}(a)]. This is a high-energy state with $E=UN^2/2$ and the left/right traps are not entangled. With Eq. (\ref{ham00}), the unitary evolution density operator reads $\rho(t)=e^{-iH_{{\rm BJJ}}t} \rho_0 e^{iH_{{\rm BJJ}}t}$. We define the reduced density matrix ${\rho_L(t)} = \text{Tr}_{{R}} [\rho(t)] $ with respect to the right well by summing over all possible states in the right well. This gives an $(N+1)\times (N+1)$ matrix with elements ${\rho}^{L}_{nn'}(t)=\sum_{k=0}^N\langle n,k|\rho(t)|n',k\rangle$. Only $k=N-n'=N-n$ survive in the summation due to particle number conservation. This implies a diagonal reduced density matrix
\begin{equation}
\begin{array}{c}
\rho_L(t) = \text{Tr}_{{R}} \, \rho(t) =
\left[
\begin{array}{c c c c c}
\rho_0^L(t) &   &     \\
   &    \ddots  &  \\ 
 &    &     \rho_N^L(t)
\end{array}
\right]
\end{array},
\label{eq3}
\end{equation}
with elements $\rho_n^L(t) = \langle n,N-n| \rho(t)|n,N-n\rangle $. For $t=0$, the $\rho(t)$ and $\rho_L(t)$ both have only a single nonzero matrix element, namely $\rho_{0,N;0,N}(t=0)=\rho_0^L(t=0)=1$.
Localization is characterized by a large weight of only a few matrix elements for all $t>0$, while for
a delocalized state the weight is distributed over time $t>0$ to all states, such that $\rho^L_k(t)\approx 1/(N+1)$.

With $\rho_L(t)$, we introduce the R\'enyi entropy~\cite{PhysRevX.8.041019} as a quantitative measure for the entanglement between the traps, which gives
\begin{equation}
	{\cal S}_\alpha (t)=\frac{1}{1-\alpha}\log_2 \text{Tr}[{\rho_L^\alpha(t)}] = \frac{1}{1-\alpha}\log_2 \sum_n  [\rho_n^L(t)]^\alpha  .
\label{eq4}
\end{equation}
In general, $\alpha$ is a free parameter and typical values used are $\alpha=2,3$~\cite{PhysRevX.8.041019}. We fix $\alpha=2$ in the subsequent calculations. ${\cal S}(t)\equiv {\cal S}_2 (t)$ measures the entanglement entropy 
over time with the initial value ${\cal S} (t=0)=0$. With Eqs. (\ref{eq3}) and (\ref{eq4}), we plot the entanglement entropy versus time for $N=100$ bosons with $J=1$ and $U=0.01 \ \text{or} \ 0.1 $ separately in Fig. \ref{fig:1}. The unitary evolution drives 
the system into higher entangled states rapidly, indicating the fast entanglement growth at the beginning. 
In the long time limit, the mean entanglement entropy saturates but is subject to strong fluctuations. The mean 
value depends on the system parameters. For instance, with $U=0.01$ ($U=0.1$) we obtain ${\cal S} \sim 3.8$ 
(${\cal S} \sim 1.5$). 

%in the limit $\alpha \rightarrow 1$, the R\'enyi entanglement entropy reproduces the Von Neumann entropy.

\begin{figure*}
    \centering
    \includegraphics[width=\linewidth]{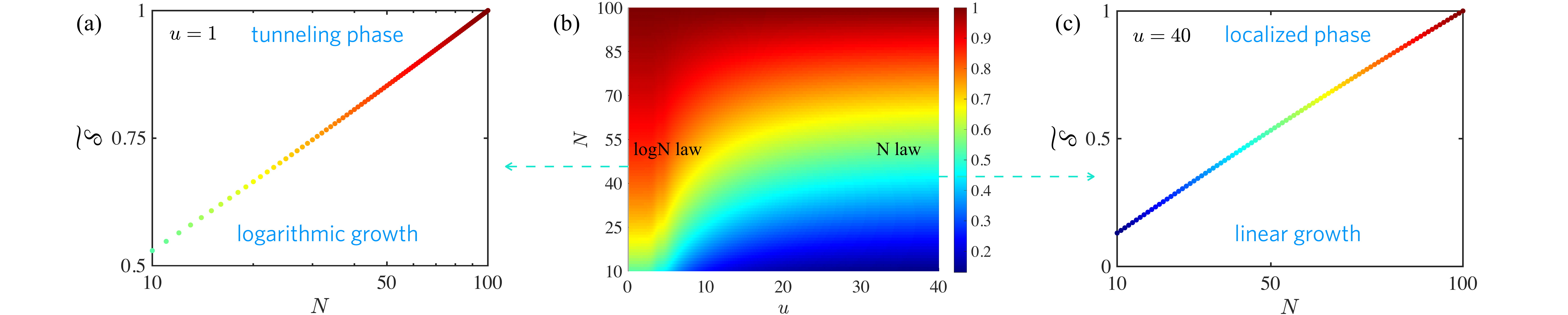}
    \caption{Scaling laws for the tunneling and HSL phases. In (a), we plot normalized entanglement entropy versus the number of bosons $N$ with $u=1$. Here $u < u_c \sim 3.7$, corresponding to the tunneling phase, ${\cal  \widetilde S}$ grows logarithmically with $N$. Different scaling behaviors are found when the system is in the HSL phase, where ${\cal  \widetilde S}$ increases linearly with $N$ (c). We plot ${\cal  \widetilde S}(u, N)$ in (b) with color representing its magnitude. The scaling law keeps ${\cal  \widetilde S} \sim \log(N)$ when $u<u_c$ and gradually becomes ${\cal  \widetilde S} \sim N $ after crossing the transition point $u_c$.}    
    \label{fig:4}
\end{figure*}

\section{Random observation time}

To reveal the generic entanglement dependence on the system parameters like $N$, $J$, and $U$, we need to avoid large quantum fluctuations. 
This is done by the observation time average method: We conduct a specific experiment (or perform a calculation) repeatedly 
at different times and average with respect to those times. In other words, we perform a time average of the observables.
This is applied to the reduced density matrix of the BJJ and gives
\begin{equation}
	\langle\rho_n^L(t)\rangle_t=\sum_{j,j'=0}^Nc_n(E_j,E_{j'})\langle e^{-i(E_j-E_{j'})t}\rangle_t,
\end{equation}
where the matrix element $c_n(E_j,E_{j'})=\langle N-n,n|E_j\rangle\langle E_j|\rho_0|E_{j'}\rangle\langle E_{j'}|N-n,n\rangle$, where $|E_j\rangle$ ($E_j$) is an eigenstate (eigenvalue) of the Hamiltonian $H_{\rm BJJ}$.
For the observation time average $\langle \cdots \rangle_t$ we choose the exponential distribution $se^{-st}dt$, where long time observations are exponentially suppressed on the scale $1/s$. This gives $ \langle e^{-i(E_j-E_{j'})t}\rangle_t=1/[1+i(E_j-E_{j'})/s]$, which indicates that the effective evolution after average is not unitary anymore. The probability is preserved, though, since $\sum_n \langle\rho_n^L(t)\rangle_t=1$, reflecting the absence of losses in the closed system. Therefore, the observation time average is different from random weak measurements. With Eq. (\ref{eq3}), the observation time average gives for the diagonal reduced density matrix $\langle \rho_L(t) \rangle_t $ with elements
 \begin{equation}
 	\langle\rho_n^L(t)\rangle_t = \int_0^{\infty} \rho_n^L(t) s e^{-st} dt = \sum_{j,j'=0}^N\frac{c_n(E_j,E_{j'})}{1+i(E_j-E_{j'})/s},
 	\label{eq6}
 \end{equation}
where $s=1$ is used subsequently.
This expression is used to calculate the corresponding R\'enyi entanglement entropy ${\cal S}(J, U, N)$ with Eq. (\ref{eq4}), where the entropy is a function of $N, U, J$ without fluctuations in time. As shown in Fig. \ref{fig:2}, the time average leads to a smooth behavior of the entanglement entropy for the system parameters, where the entanglement transition by the HSL is clearly presented. 

In Fig. \ref{fig:2}(b,d), we plot ${\cal S} (J)$ for different number of bosons. Physically, when $J$ is small, i.e., in the HSL phase, the motion of the bosons is constrained to the southern hemisphere [Fig. \ref{fig:2}(a)] and the matrix element $\langle N-k,k|H_{{\rm BJJ}}|0,N\rangle=-J\langle k,N-k|a_L^\dagger a_R + a_R^\dagger a_L|0,N\rangle$ is very small and proportional to $J$ (for fixed $N$). This is exactly reflected in the entanglement entropy, where ${\cal S}(J)$ grows approximately linearly with $J$ until it reaches the critical value $J_c$ [Fig. \ref{fig:2}(b)]. The $J_c$ is the transition point of the HSL phase to the tunneling phase, after which the bosons can reach the entire Hilbert space from the initial state [Fig. \ref{fig:2}(a)]. This explains why in the tunneling regime the ${\cal S}(J)$ is a constant and only depends on the size of the system [Fig. \ref{fig:2}(d)], meaning the entanglement entropy measures to what extent the system can explore the Hilbert space. In Fig. \ref{fig:2}(c), the competition between the tunneling $J$ and particle interactions $U$ for $N=20$ is presented. When increasing $U$, it takes $J$ larger value to reach the entanglement phase transition. In the HSL phase, the ${\cal S}(J, U)$ is proportional to $J$ and inversely proportional to $U$. While in the tunneling regime ${\cal S}(J, U)$ is a constant, when the entire Hilbert space is accessible. In Fig. \ref{fig:2}(e), we plot ${\cal S}(U,N)$, where the size effects on the entanglement phase transition are shown. When the number of bosons is large, the system is more easily localized in the Hilbert space and undergoes the entanglement phase transition for 
smaller values of $U$, since the critical value $u_c=UN/J$ does not depend on N.   
The critical entanglement phase transition point is shown in Fig. \ref{fig:3} with the characteristic parameter $u=UN/J$, which reveals a sharp boundary between the tunneling and HSL phases. In Fig. \ref{fig:3}(a), we plot the ${\cal S}$ versus $J$ rescaled by the $N$. Here, $U$ is set to 1, and the entanglement entropy undergoes the phase transition, for different numbers of bosons, at the same value of $J/N \sim 0.27$. The critical point $u_c = UN/J \sim 3.7$. Similarly, we plot ${\cal S}(U,N)$ for different $N$s in Fig. \ref{fig:3}(b), where the system also exhibits the entanglement transition at $u_c \sim 3.7$. This generic phase transition behavior is also confirmed by tuning $N$ with the rescaling by $J^{-1}$ [Fig. \ref{fig:3}(c)] and $U$ [Fig. \ref{fig:3}(d)], where the entanglement reaches maximum at the $u_c \sim 3.7$. Interestingly, this quantum phase transition point deviates from the results by the Gross-Pitaevskii equation~\cite{PhysRevA.55.4318} under the mean-field approximation, where 
\begin{equation}
	 3.7 \approx u_c^{\rm Quantum} < u_c^{\rm Mean-field} = 4.
\end{equation}

It is crucial to note that both critical values depend on the number of bosons $N$ only through the combination
of parameters $u=UN/J$, as demonstrated in Fig. \ref{fig:3}.
This fact, as well as the critical value $u_c^{\rm Quantum}\approx 3.7$, were also found 
previously for the scaling change of the participation ratio~\cite{cohen16} and for the jump of the return
probability~\cite{symmetry21}. 
%Interestingly though, the critical value depends weakly on the initial conditions. 
%{\color{blue}
%The critical value is defined as the maximum of the EE, the maximum of the participation ratio and the
%maximum of the transition probability, respectively. 
%}
The above result reflects the effective enhancement of the particle-particle interaction by quantum entanglement, where the system undergoes the HSL phase transition with smaller $U$s and causes quantitative change in the entropy. In contrast, the mean-field approximation ignores the entanglement and therefore, it requires a stronger particle-particle interaction to reach the self-trapping transition. 
The deviation of $u_c$ from its mean-field value is in agreement with a recent work by Wimberger et al. on the $N$
dependence of $u_c$ for different initial
states~\cite{PhysRevA.103.023326}. Their critical values are monotonically increasing even up to $N\approx 100$,
in contrast to Fig. \ref{fig:3}. This indicates that the observation time averaged EE 
provides a more stable criterion for the definition of a critical point.

Our calculations are limited to $N\approx 100$ bosons. It would be
interesting though to calculate the critical value $u_c$ also in the large $N$ regime. The fact that the EE
increases monotonically with $N$ in Fig. \ref{fig:3} indicates that a classical description with only a one-particle
mean-field wavefunction for the two sites of the BJJ might not be sufficient, at least for $u\approx u_c$. 
A semiclassical approximation has been used as an interpolation between
the full quantum evolution and the classical approximation, where large values of $N$
are accessible~\cite{PhysRevA.82.053617,PhysRevA.103.023326}. Such calculations reveal that 
the value of $u_c$ varies for different initial states. Moreover, the fluctuations decay very slowly
with $N$ in the vicinity of $u_c$~\cite{PhysRevA.82.053617}.

Different scaling behaviors with the size of the system are found in HSL phase and tunneling phase. Here we use the normalized entanglement entropy defined by ${\cal  \widetilde S}  = {\cal S}/{\cal S}_{max}$ for constant $u$. As shown in Figs. \ref{fig:4}(a,c), we plot ${\cal  \widetilde S}$ versus the number of the bosons $N$ keeping characteristic parameter $u$ is a constant [$u=1<u_c$ for Fig. \ref{fig:4}(a) and $u=40>u_c$ for Fig. \ref{fig:4}(c)]. Note as $u \sim J^{-1}$ and $u \sim UN$, the small $u$ corresponds to the tunneling phase and large $u$ corresponds to HSL phase. The entanglement entropy increases logarithmically  with the size of the system for the tunneling regime and linearly for the localized regime, namely
\begin{equation}
	{\cal  \widetilde S} \sim \log(N), \ u < u_c ; \ \ \ \  \ {\cal  \widetilde S} \sim N, \ u>u_c.
\end{equation}
The linear $N$ behavior is counter-intuitive when we follow the argument of a single localized state. In the present 
model, though, localization appears in a two-dimensional space that includes two Fock states. This is a consequence
of the mirror symmetry of the BJJ model. Then the  EE reads 
${\cal  \widetilde S} \sim -\log_2[\rho_N^2+(1-\rho_N)^2]\sim 2\rho_N$ with $\rho_N\sim c N$.
This was confirmed in a direct calculation of the reduced density matrix with $c\approx 0.002$ 
(cf. App. \ref{app:evolution}). Then the linear $N$ behavior is valid in a crossover region up to $N\approx 200$
but becomes asymptotically $\rho_N\sim 1/2$.
The transition of scaling behaviors around $u_c \sim 3.7 $ is clearly visualized in Fig. \ref{fig:4}(b).  The linear increase reflects the fact that a fixed fraction of states is localized, and the entanglement entropy increases with the effective volume of the traced right well. This is obviously quite different from the tunneling regime, where the entire Hilbert space is involved, implying that the matrix elements of the reduced density matrix are equally distributed, leading to a $\log N$ behavior. This logarithmic behavior can be considered a modification of the area-law, which is initially proposed for non-interacting particles.

\begin{figure}
    \centering
    \includegraphics[width=\linewidth]{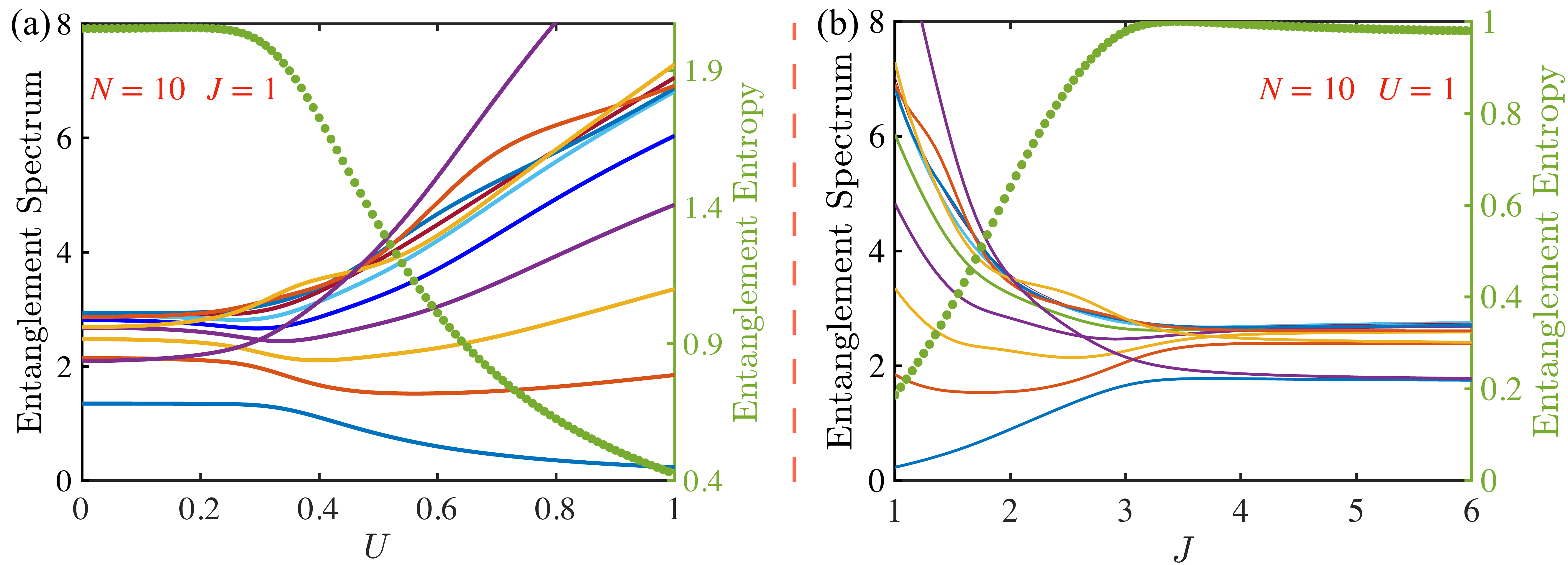}
    \caption{Entanglement spectrum as a function of $U$ (a) and $J$ (b). The transition to the HSL is indicated by the repulsion of the levels, which keeps constant in the tunneling phase and spread in the HSL phase. Compare with the entanglement entropy, the critical transition of the entanglement spectrum is again characterized by $u_c$.}
     \label{fig5}
\end{figure}

Quite generally, a degeneracy in the energy spectrum can be associated with a phase transition. For instance, a typical quantum phase transition is associated with the degeneracy of the ground state. Then an important question is: whether the above entanglement phase transition reveals some characteristic features in the entanglement spectrum~\cite{PhysRevLett.101.010504},  which is defined as the eigenvalues of the reduced density matrix. The concept of entanglement spectrum is utilized in the detecting of topological orders~\cite{PhysRevLett.121.250601} and investigated extensively with different lattice models~\cite{PhysRevLett.121.200602,PhysRevLett.123.211603}. In the case of the evolution with excited states, the relation between the phase transition and the entanglement spectrum can be very complex because, as in the case of HSL, it is not a transition associated with the ground state alone. Here, the diagonal form of the reduced density matrix in Eq. (\ref{eq3}) enables the direct calculation of the entanglement spectrum. Together with Eq. (\ref{eq6}), we have
\begin{equation}
	\xi_n = \log [ \langle\rho_n^L(t)\rangle_t].
\end{equation}
Here we use the time-independent reduced matrix to avoid the fluctuations as we have discussed.
We plot the entanglement spectrum $\xi_n(U)$ in Fig. \ref{fig5}(a) and $\xi_n(J)$ in Fig. \ref{fig5}(b) for $N=10$, where the entanglement transition exhibits clearly signatures in the spectrum. Specifically, the entanglement spectrum has almost constant levels independent of $U$ and $J$ in the tunneling phase and repulsive levels in the HSL phase. The transition from constant levels to repulsive levels happens exactly at the critical transition point $u_c$ as we find for the entanglement entropy. Thus, the entanglement spectrum also is a useful indicator for observing a qualitative change in the onset of the HSL transition.
In particular, the strong level repulsion in the localized regime with increasing $U$ is significant.

%{\it Summary and discussion.---}
\section{Summary and discussion}

We have investigated the entanglement phase transition that arises from the competition between the tunneling and particle-particle interaction, where both entanglement entropy and entanglement spectrum undergo a clear quantum phase transition from the tunneling regime to the HSL regime. It is crucial though to apply the observation time average to suppress strong quantum fluctuations in time, which is a powerful tool and can be extended for other studies on entanglement. Moreover, the transition point we find here is smaller than the value given by the mean-field approximation with the Gross-Pitaevskii equation. This reveals the quantum effects that are not covered in the form of self-trapping. Different scaling laws with the size of the system are found before and after the phase transition. It is shown that the entanglement entropy grows logarithmically with the number of bosons in the tunneling phase and linearly in the localized phase. 

Our results indicate that the entanglement entropy (entanglement spectrum) can be controlled by the tunneling rate $J$, interaction strength $U$, and the number of the bosons $N$, which can be proven in the present experiment platforms. For instance, for an ultracold Bose gas in a two-site optical lattice, the potential barrier can be tuned by an external Laser field~\cite{PhysRevLett.120.173601}. Another realization of a BJJ is a pair of coupled polariton condensates~\cite{Abbarchi2013}. In such experiments, by tuning the parameters in the setup, critical phase transitions may be observed. The generalization of the two-site structure of the BJJ to more sites should provide similar effects of controllable entanglement and HSL, reflecting that the two-site prototype is a scalable model. Another extension of the BJJ is the coupling to a large but finite bath~\cite{PRXQuantum.2.010340}. Then we expect similar localization and entanglement effects as we found here.

\begin{acknowledgments}
We are grateful to Eli Barkai for the useful discussions. Q.L. thanks Qiming Ding, and Qiang Miao for illuminating discussions. This research is supported by the Israel Science Foundation through Grant No. 1614/21 (Q.L.) and by a grant of the Julian Schwinger Foundation for Physics Research (K.Z.).	
\end{acknowledgments}

\appendix

\section{Evolution of the entanglement entropy
}
\label{app:evolution}

\begin{figure*}
    \centering
    \includegraphics[width=16cm,height=8.5cm]{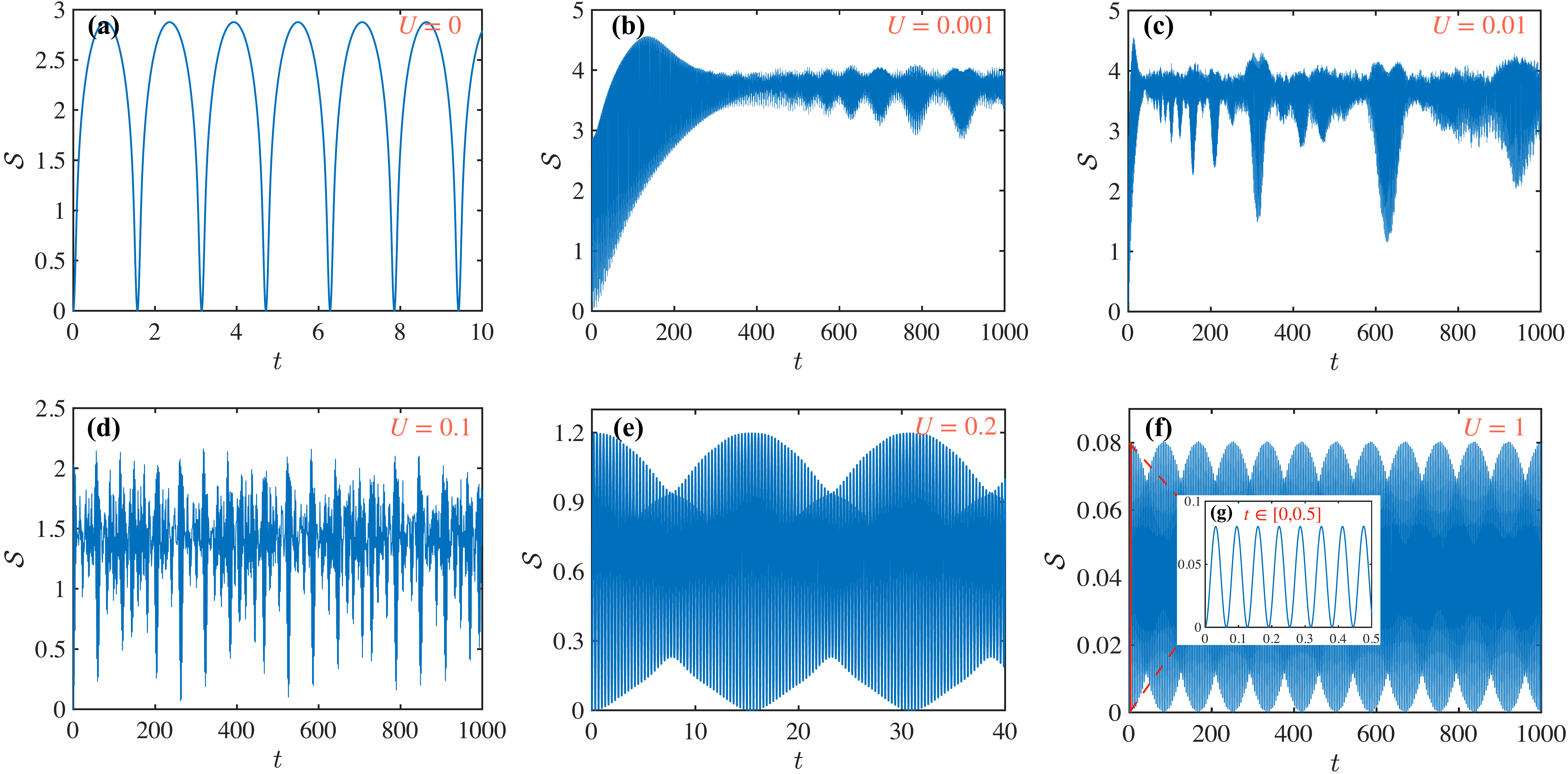}%{68sup1.eps} \\ %{EETimeEv.eps}\\
    \caption{Evolution of the EE for $N=100$ bosons with $J=1$ and different values of the interaction strength $U$
    on different time scales.}    
    \label{fig:11}
\end{figure*}

The EE is given by the reduced density matrix ${\hat\rho}$ through the R\'enyi entropy. Thus, the evolution of the EE
is determined by the evolution of ${\hat\rho}$, whose diagonal elements can be written in spectral representation as
\[
{\hat\rho}_n(t)=\langle n,N-n|e^{-iHt}|0,N\rangle\langle0,N|e^{iHt}|n,N-n\rangle
\]
\[
=\sum_{j,j'=0}^N e^{-i(E_j-E_{j'})t}
\]
\begin{equation}
\label{expansion}
\langle n,N-n|E_j\rangle\langle E_j|0,N\rangle\langle0,N|E_{j'}\rangle\langle E_{j'}|n,N-n\rangle
\ .
\end{equation}
This is a superposition of oscillating functions with frequencies $\{|E_j-E_{j'}|\}$, where the fastest oscillations are
given by the largest frequency.  We can distinguish two extreme cases for the BJJ, the non-interacting case with
$U=0$ and the non-tunneling case with $J=0$. For $U=0$ the energy spectrum is equidistant with distance $2J$
and energy eigenvalues $E_0=0$, $E_j=\pm 2Jj$ ($j=1,2,...,N/2$) \cite{Ziegler2012}.
This implies a periodic evolution with period $t_p=\pi/2J$. In particular, the matrix elements of the density 
matrix elements with $n=0, N$ read ${\hat\rho}_{0}(t)=|\cos(Jt)|^{2N}$ and 
${\hat\rho}_{N}(t)=|\sin(Jt)|^{2N}$ \cite{Ziegler2017}.
This is reflected by the behavior visualized in Fig. \ref{fig:11}a.
The other extreme is $J=0$. Since the initial state $|0,N\rangle$ is an eigenstate of the Hamiltonian, there is no
evolution. A small tunneling rate $J$, though, splits the degeneracy of $|0,N\rangle$ and $|N,0\rangle$ and creates a small
frequency of order $J$ that leads to very slow oscillations in the evolution.
On the other hand, the fastest oscillations are caused by the maximal frequency $UN^2/4$. 
This is visible in Fig. \ref{fig:11}f.

\begin{figure}
    \centering
    \includegraphics[width=7cm,height=5cm]{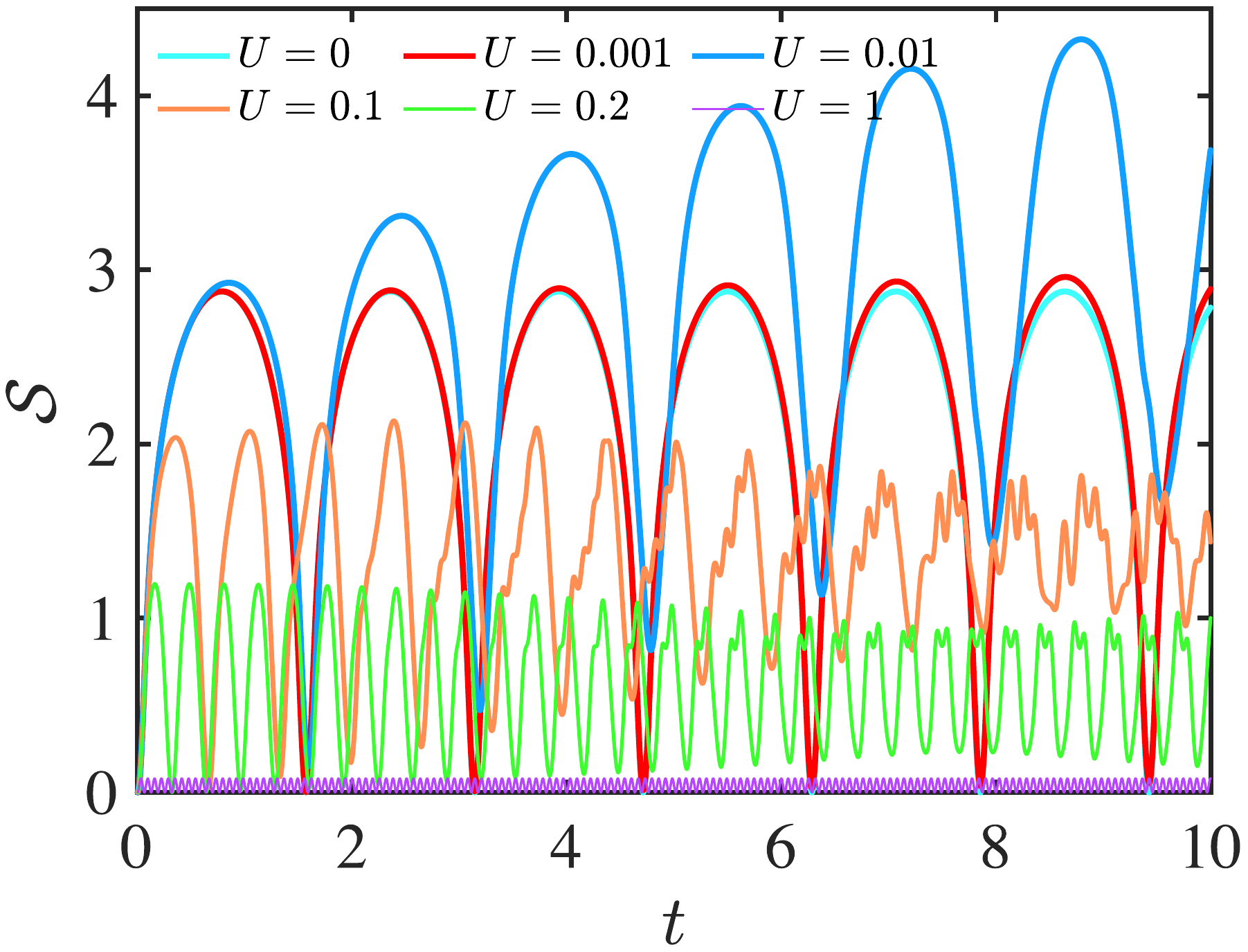} \\ %{EETimeEv.eps}\\
    \caption{Evolution of the EE for different values of $U$, $J=1$ and $N=100$ on the same time scale.}    
    \label{fig:12}
\end{figure}

The evolution is more complex in the intermediate regime, where tunneling and interaction compete.
Different frequencies contribute and there is an evolution on all time scales, caused by frequencies 
from 0 to the largest frequency $UN^2/4$. The oscillating behavior is only limited 
by the number of frequencies $N_f =(N+1)N/2$.
Typical examples of the EE in the BJJ are plotted in Fig.\ref{fig:11} on different time scales. The influence of the interaction
strength $U$ distinguishes clearly a non-interacting tunneling regime (Fig. \ref{fig:11}a) and a strongly interacting
regime of fast oscillations with $t_p\approx0.07$ (inset in Fig. \ref{fig:11}f) and
slower oscillations with $t'_p\approx 90$, where the EE is quite small.
There is an intermediate regime that has characteristic features on long time scales that represents small frequencies.
Moreover, the intermediate regime has a higher EE than the other two regimes. This indicates a more complex
evolution.  In Fig. \ref{fig:12} the evolution of the EE is compared on the same time scale for the different regimes. 

\begin{figure*}
    \centering
    \includegraphics[width=12cm,height=5cm]{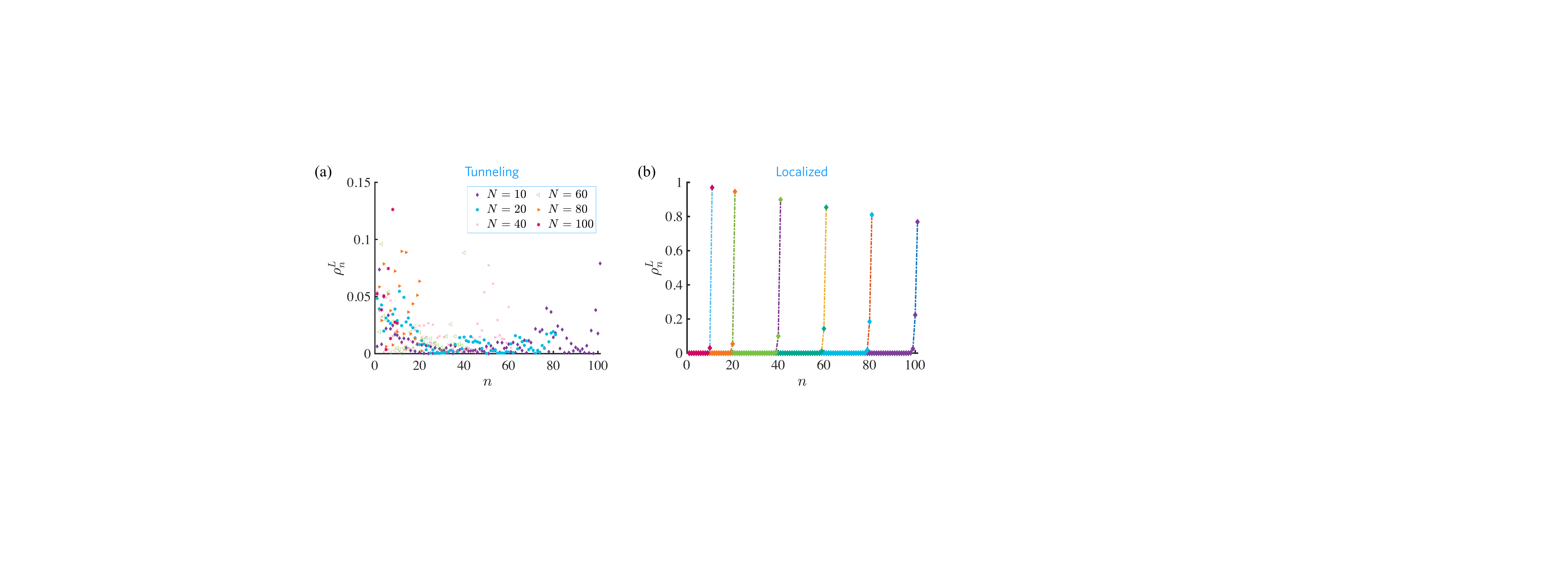}
    \caption{Maximum of the elements $\rho^L_n(t)$ of the reduced density matrix on the time interval 
    $[0,2000]$: a) Deep in the tunneling regime ($u=1$). 
    b) Deep in the localized regime ($u=40$) for different numbers of bosons $N$.}    
    \label{fig:13}
\end{figure*}

Besides the frequencies of the oscillations, the expansion coefficients in (\ref{expansion}) play a crucial role in the
evolution. They reflect the overlap between the Fock states and the energy eigenstates and determine the entanglement
of the different Fock states. 
In the tunneling regime all elements of the reduced density matrix contribute significantly, for instance, at time $t=1000$ 
(cf. Fig. \ref{fig:13}a). 
In the localized regime, on the other hand, the overlaps between the Fock states with a few energy
eigenstates are large. In particular, for $J=0$ we get ${\hat\rho}_n(t)=\delta_{nN}$, which yields a vanishing EE.
Small tunneling can be treated as a perturbation expansion in powers of $J$, leading to a small EE. In this case
the elements of the reduced density matrix have only two significant contributions, as visualized in Fig. \ref{fig:13}b.
A special effect is that for a fixed $u$ the smaller of the two matrix elements increases linearly with $N$ like $2N/1000$ 
up to $N\approx200$ bosons. We anticipate a saturation for larger values of $N$, which results eventually in two equal
matrix elements of $1/2$.   

These time-dependent results indicate a complex evolution of the reduced density matrix and the EE that is 
caused by the oscillating
behavior with various frequencies. Degeneracies in the energy spectrum due to specific physical effects, such as phase
transitions, appear on large time scales. This fact suggests that the oscillating behavior of the reduced density matrix
on short time scales should be averaged out.  

%%%%%%%%%%%%%%%%%%%%%%%%%%%%%%%%%%%%%%%%%%%%%%%
%%%%%%%%%%%%%%%%%%%%%%%%%%%%%%%%%%%%%%%%%%%%%%

%\bibliography{ref}

\begin{thebibliography}{99}

\bibitem{Ziegler2012}
 K. Ziegler, Laser Physics {\bf 22}, 331 (2012)

\bibitem{Ziegler2017}
K. Ziegler, International Journal of Modern Physics B {\bf 31}, 1750255 (2017)

\end{thebibliography}

\begin{thebibliography}{48}%
\makeatletter
\providecommand \@ifxundefined [1]{%
 \@ifx{#1\undefined}
}%
\providecommand \@ifnum [1]{%
 \ifnum #1\expandafter \@firstoftwo
 \else \expandafter \@secondoftwo
 \fi
}%
\providecommand \@ifx [1]{%
 \ifx #1\expandafter \@firstoftwo
 \else \expandafter \@secondoftwo
 \fi
}%
\providecommand \natexlab [1]{#1}%
\providecommand \enquote  [1]{``#1''}%
\providecommand \bibnamefont  [1]{#1}%
\providecommand \bibfnamefont [1]{#1}%
\providecommand \citenamefont [1]{#1}%
\providecommand \href@noop [0]{\@secondoftwo}%
\providecommand \href [0]{\begingroup \@sanitize@url \@href}%
\providecommand \@href[1]{\@@startlink{#1}\@@href}%
\providecommand \@@href[1]{\endgroup#1\@@endlink}%
\providecommand \@sanitize@url [0]{\catcode `\\12\catcode `\$12\catcode
  `\&12\catcode `\#12\catcode `\^12\catcode `\_12\catcode `\%12\relax}%
\providecommand \@@startlink[1]{}%
\providecommand \@@endlink[0]{}%
\providecommand \url  [0]{\begingroup\@sanitize@url \@url }%
\providecommand \@url [1]{\endgroup\@href {#1}{\urlprefix }}%
\providecommand \urlprefix  [0]{URL }%
\providecommand \Eprint [0]{\href }%
\providecommand \doibase [0]{http://dx.doi.org/}%
\providecommand \selectlanguage [0]{\@gobble}%
\providecommand \bibinfo  [0]{\@secondoftwo}%
\providecommand \bibfield  [0]{\@secondoftwo}%
\providecommand \translation [1]{[#1]}%
\providecommand \BibitemOpen [0]{}%
\providecommand \bibitemStop [0]{}%
\providecommand \bibitemNoStop [0]{.\EOS\space}%
\providecommand \EOS [0]{\spacefactor3000\relax}%
\providecommand \BibitemShut  [1]{\csname bibitem#1\endcsname}%
\let\auto@bib@innerbib\@empty
%</preamble>
\bibitem [{\citenamefont {Bombelli}\ \emph {et~al.}(1986)\citenamefont
  {Bombelli}, \citenamefont {Koul}, \citenamefont {Lee},\ and\ \citenamefont
  {Sorkin}}]{PhysRevD.34.373}%
  \BibitemOpen
  \bibfield  {author} {\bibinfo {author} {\bibfnamefont {L.}~\bibnamefont
  {Bombelli}}, \bibinfo {author} {\bibfnamefont {R.~K.}\ \bibnamefont {Koul}},
  \bibinfo {author} {\bibfnamefont {J.}~\bibnamefont {Lee}}, \ and\ \bibinfo
  {author} {\bibfnamefont {R.~D.}\ \bibnamefont {Sorkin}},\ }\href {\doibase
  10.1103/PhysRevD.34.373} {\bibfield  {journal} {\bibinfo  {journal} {Phys.
  Rev. D}\ }\textbf {\bibinfo {volume} {34}},\ \bibinfo {pages} {373} (\bibinfo
  {year} {1986})}\BibitemShut {NoStop}%
\bibitem [{\citenamefont {Srednicki}(1993)}]{PhysRevLett.71.666}%
  \BibitemOpen
  \bibfield  {author} {\bibinfo {author} {\bibfnamefont {M.}~\bibnamefont
  {Srednicki}},\ }\href {\doibase 10.1103/PhysRevLett.71.666} {\bibfield
  {journal} {\bibinfo  {journal} {Phys. Rev. Lett.}\ }\textbf {\bibinfo
  {volume} {71}},\ \bibinfo {pages} {666} (\bibinfo {year} {1993})}\BibitemShut
  {NoStop}%
\bibitem [{\citenamefont {Eisert}\ \emph {et~al.}(2010)\citenamefont {Eisert},
  \citenamefont {Cramer},\ and\ \citenamefont {Plenio}}]{RevModPhys.82.277}%
  \BibitemOpen
  \bibfield  {author} {\bibinfo {author} {\bibfnamefont {J.}~\bibnamefont
  {Eisert}}, \bibinfo {author} {\bibfnamefont {M.}~\bibnamefont {Cramer}}, \
  and\ \bibinfo {author} {\bibfnamefont {M.~B.}\ \bibnamefont {Plenio}},\
  }\href {\doibase 10.1103/RevModPhys.82.277} {\bibfield  {journal} {\bibinfo
  {journal} {Rev. Mod. Phys.}\ }\textbf {\bibinfo {volume} {82}},\ \bibinfo
  {pages} {277} (\bibinfo {year} {2010})}\BibitemShut {NoStop}%
\bibitem [{\citenamefont {Miao}\ and\ \citenamefont
  {Barthel}(2021)}]{PhysRevLett.127.040603}%
  \BibitemOpen
  \bibfield  {author} {\bibinfo {author} {\bibfnamefont {Q.}~\bibnamefont
  {Miao}}\ and\ \bibinfo {author} {\bibfnamefont {T.}~\bibnamefont {Barthel}},\
  }\href {\doibase 10.1103/PhysRevLett.127.040603} {\bibfield  {journal}
  {\bibinfo  {journal} {Phys. Rev. Lett.}\ }\textbf {\bibinfo {volume} {127}},\
  \bibinfo {pages} {040603} (\bibinfo {year} {2021})}\BibitemShut {NoStop}%
\bibitem [{\citenamefont {Li}\ \emph {et~al.}(2019)\citenamefont {Li},
  \citenamefont {Chen},\ and\ \citenamefont {Fisher}}]{PhysRevB.100.134306}%
  \BibitemOpen
  \bibfield  {author} {\bibinfo {author} {\bibfnamefont {Y.}~\bibnamefont
  {Li}}, \bibinfo {author} {\bibfnamefont {X.}~\bibnamefont {Chen}}, \ and\
  \bibinfo {author} {\bibfnamefont {M.~P.~A.}\ \bibnamefont {Fisher}},\ }\href
  {\doibase 10.1103/PhysRevB.100.134306} {\bibfield  {journal} {\bibinfo
  {journal} {Phys. Rev. B}\ }\textbf {\bibinfo {volume} {100}},\ \bibinfo
  {pages} {134306} (\bibinfo {year} {2019})}\BibitemShut {NoStop}%
\bibitem [{\citenamefont {Skinner}\ \emph {et~al.}(2019)\citenamefont
  {Skinner}, \citenamefont {Ruhman},\ and\ \citenamefont
  {Nahum}}]{PhysRevX.9.031009}%
  \BibitemOpen
  \bibfield  {author} {\bibinfo {author} {\bibfnamefont {B.}~\bibnamefont
  {Skinner}}, \bibinfo {author} {\bibfnamefont {J.}~\bibnamefont {Ruhman}}, \
  and\ \bibinfo {author} {\bibfnamefont {A.}~\bibnamefont {Nahum}},\ }\href
  {\doibase 10.1103/PhysRevX.9.031009} {\bibfield  {journal} {\bibinfo
  {journal} {Phys. Rev. X}\ }\textbf {\bibinfo {volume} {9}},\ \bibinfo {pages}
  {031009} (\bibinfo {year} {2019})}\BibitemShut {NoStop}%
\bibitem [{\citenamefont {Lunt}\ \emph {et~al.}(2021)\citenamefont {Lunt},
  \citenamefont {Szyniszewski},\ and\ \citenamefont
  {Pal}}]{PhysRevB.104.155111}%
  \BibitemOpen
  \bibfield  {author} {\bibinfo {author} {\bibfnamefont {O.}~\bibnamefont
  {Lunt}}, \bibinfo {author} {\bibfnamefont {M.}~\bibnamefont {Szyniszewski}},
  \ and\ \bibinfo {author} {\bibfnamefont {A.}~\bibnamefont {Pal}},\ }\href
  {\doibase 10.1103/PhysRevB.104.155111} {\bibfield  {journal} {\bibinfo
  {journal} {Phys. Rev. B}\ }\textbf {\bibinfo {volume} {104}},\ \bibinfo
  {pages} {155111} (\bibinfo {year} {2021})}\BibitemShut {NoStop}%
\bibitem [{\citenamefont {Amico}\ \emph {et~al.}(2008)\citenamefont {Amico},
  \citenamefont {Fazio}, \citenamefont {Osterloh},\ and\ \citenamefont
  {Vedral}}]{amicoRevModPhys.80.517}%
  \BibitemOpen
  \bibfield  {author} {\bibinfo {author} {\bibfnamefont {L.}~\bibnamefont
  {Amico}}, \bibinfo {author} {\bibfnamefont {R.}~\bibnamefont {Fazio}},
  \bibinfo {author} {\bibfnamefont {A.}~\bibnamefont {Osterloh}}, \ and\
  \bibinfo {author} {\bibfnamefont {V.}~\bibnamefont {Vedral}},\ }\href
  {\doibase 10.1103/RevModPhys.80.517} {\bibfield  {journal} {\bibinfo
  {journal} {Rev. Mod. Phys.}\ }\textbf {\bibinfo {volume} {80}},\ \bibinfo
  {pages} {517} (\bibinfo {year} {2008})}\BibitemShut {NoStop}%
\bibitem [{\citenamefont {Abanin}\ \emph {et~al.}(2019)\citenamefont {Abanin},
  \citenamefont {Altman}, \citenamefont {Bloch},\ and\ \citenamefont
  {Serbyn}}]{RevModPhys.91.021001}%
  \BibitemOpen
  \bibfield  {author} {\bibinfo {author} {\bibfnamefont {D.~A.}\ \bibnamefont
  {Abanin}}, \bibinfo {author} {\bibfnamefont {E.}~\bibnamefont {Altman}},
  \bibinfo {author} {\bibfnamefont {I.}~\bibnamefont {Bloch}}, \ and\ \bibinfo
  {author} {\bibfnamefont {M.}~\bibnamefont {Serbyn}},\ }\href {\doibase
  10.1103/RevModPhys.91.021001} {\bibfield  {journal} {\bibinfo  {journal}
  {Rev. Mod. Phys.}\ }\textbf {\bibinfo {volume} {91}},\ \bibinfo {pages}
  {021001} (\bibinfo {year} {2019})}\BibitemShut {NoStop}%
\bibitem [{\citenamefont {Vidal}\ \emph {et~al.}(2003)\citenamefont {Vidal},
  \citenamefont {Latorre}, \citenamefont {Rico},\ and\ \citenamefont
  {Kitaev}}]{kitaevPhysRevLett.90.227902}%
  \BibitemOpen
  \bibfield  {author} {\bibinfo {author} {\bibfnamefont {G.}~\bibnamefont
  {Vidal}}, \bibinfo {author} {\bibfnamefont {J.~I.}\ \bibnamefont {Latorre}},
  \bibinfo {author} {\bibfnamefont {E.}~\bibnamefont {Rico}}, \ and\ \bibinfo
  {author} {\bibfnamefont {A.}~\bibnamefont {Kitaev}},\ }\href {\doibase
  10.1103/PhysRevLett.90.227902} {\bibfield  {journal} {\bibinfo  {journal}
  {Phys. Rev. Lett.}\ }\textbf {\bibinfo {volume} {90}},\ \bibinfo {pages}
  {227902} (\bibinfo {year} {2003})}\BibitemShut {NoStop}%
\bibitem [{\citenamefont {Calabrese}\ and\ \citenamefont
  {Cardy}(2004)}]{calabrese2004}%
  \BibitemOpen
  \bibfield  {author} {\bibinfo {author} {\bibfnamefont {P.}~\bibnamefont
  {Calabrese}}\ and\ \bibinfo {author} {\bibfnamefont {J.}~\bibnamefont
  {Cardy}},\ }\href {\doibase 10.1088/1742-5468/2004/06/p06002} {\bibfield
  {journal} {\bibinfo  {journal} {J. Stat. Phys Mech.}\ }\textbf {\bibinfo
  {volume} {2004}},\ \bibinfo {pages} {P06002} (\bibinfo {year}
  {2004})}\BibitemShut {NoStop}%
\bibitem [{\citenamefont {Peschel}\ and\ \citenamefont
  {Eisler}(2009)}]{peschel2009}%
  \BibitemOpen
  \bibfield  {author} {\bibinfo {author} {\bibfnamefont {I.}~\bibnamefont
  {Peschel}}\ and\ \bibinfo {author} {\bibfnamefont {V.}~\bibnamefont
  {Eisler}},\ }\href {\doibase 10.1088/1751-8113/42/50/504003} {\bibfield
  {journal} {\bibinfo  {journal} {J. Phys. A: Math. Theor.}\ }\textbf {\bibinfo
  {volume} {42}},\ \bibinfo {pages} {504003} (\bibinfo {year}
  {2009})}\BibitemShut {NoStop}%
\bibitem [{\citenamefont {Calabrese}\ and\ \citenamefont
  {Cardy}(2009)}]{calabrese2009}%
  \BibitemOpen
  \bibfield  {author} {\bibinfo {author} {\bibfnamefont {P.}~\bibnamefont
  {Calabrese}}\ and\ \bibinfo {author} {\bibfnamefont {J.}~\bibnamefont
  {Cardy}},\ }\href {\doibase 10.1088/1751-8113/42/50/504005} {\bibfield
  {journal} {\bibinfo  {journal} {J. Phys. A: Math. Theor.}\ }\textbf {\bibinfo
  {volume} {42}},\ \bibinfo {pages} {504005} (\bibinfo {year}
  {2009})}\BibitemShut {NoStop}%
\bibitem [{\citenamefont {Serbyn}\ \emph {et~al.}(2015)\citenamefont {Serbyn},
  \citenamefont {Papi\ifmmode~\acute{c}\else \'{c}\fi{}},\ and\ \citenamefont
  {Abanin}}]{PhysRevX.5.041047}%
  \BibitemOpen
  \bibfield  {author} {\bibinfo {author} {\bibfnamefont {M.}~\bibnamefont
  {Serbyn}}, \bibinfo {author} {\bibfnamefont {Z.}~\bibnamefont
  {Papi\ifmmode~\acute{c}\else \'{c}\fi{}}}, \ and\ \bibinfo {author}
  {\bibfnamefont {D.~A.}\ \bibnamefont {Abanin}},\ }\href {\doibase
  10.1103/PhysRevX.5.041047} {\bibfield  {journal} {\bibinfo  {journal} {Phys.
  Rev. X}\ }\textbf {\bibinfo {volume} {5}},\ \bibinfo {pages} {041047}
  (\bibinfo {year} {2015})}\BibitemShut {NoStop}%
\bibitem [{\citenamefont {Kos}\ \emph {et~al.}(2018)\citenamefont {Kos},
  \citenamefont {Ljubotina},\ and\ \citenamefont {Prosen}}]{PhysRevX.8.021062}%
  \BibitemOpen
  \bibfield  {author} {\bibinfo {author} {\bibfnamefont {P.}~\bibnamefont
  {Kos}}, \bibinfo {author} {\bibfnamefont {M.}~\bibnamefont {Ljubotina}}, \
  and\ \bibinfo {author} {\bibfnamefont {T.~c.~v.}\ \bibnamefont {Prosen}},\
  }\href {\doibase 10.1103/PhysRevX.8.021062} {\bibfield  {journal} {\bibinfo
  {journal} {Phys. Rev. X}\ }\textbf {\bibinfo {volume} {8}},\ \bibinfo {pages}
  {021062} (\bibinfo {year} {2018})}\BibitemShut {NoStop}%
\bibitem [{\citenamefont {Chan}\ \emph {et~al.}(2018)\citenamefont {Chan},
  \citenamefont {De~Luca},\ and\ \citenamefont {Chalker}}]{PhysRevX.8.041019}%
  \BibitemOpen
  \bibfield  {author} {\bibinfo {author} {\bibfnamefont {A.}~\bibnamefont
  {Chan}}, \bibinfo {author} {\bibfnamefont {A.}~\bibnamefont {De~Luca}}, \
  and\ \bibinfo {author} {\bibfnamefont {J.~T.}\ \bibnamefont {Chalker}},\
  }\href {\doibase 10.1103/PhysRevX.8.041019} {\bibfield  {journal} {\bibinfo
  {journal} {Phys. Rev. X}\ }\textbf {\bibinfo {volume} {8}},\ \bibinfo {pages}
  {041019} (\bibinfo {year} {2018})}\BibitemShut {NoStop}%
\bibitem [{\citenamefont {Nandkishore}\ and\ \citenamefont
  {Huse}(2015)}]{doi:10.1146/annurev-conmatphys-031214-014726}%
  \BibitemOpen
  \bibfield  {author} {\bibinfo {author} {\bibfnamefont {R.}~\bibnamefont
  {Nandkishore}}\ and\ \bibinfo {author} {\bibfnamefont {D.~A.}\ \bibnamefont
  {Huse}},\ }\href {https://doi.org/10.1146/annurev-conmatphys-031214-014726}
  {\bibfield  {journal} {\bibinfo  {journal} {Annu. Rev. Condens. Matter
  Phys.}\ }\textbf {\bibinfo {volume} {6}},\ \bibinfo {pages} {15} (\bibinfo
  {year} {2015})}\BibitemShut {NoStop}%
\bibitem [{\citenamefont {Levin}\ and\ \citenamefont
  {Wen}(2006)}]{PhysRevLett.96.110405}%
  \BibitemOpen
  \bibfield  {author} {\bibinfo {author} {\bibfnamefont {M.}~\bibnamefont
  {Levin}}\ and\ \bibinfo {author} {\bibfnamefont {X.-G.}\ \bibnamefont
  {Wen}},\ }\href {\doibase 10.1103/PhysRevLett.96.110405} {\bibfield
  {journal} {\bibinfo  {journal} {Phys. Rev. Lett.}\ }\textbf {\bibinfo
  {volume} {96}},\ \bibinfo {pages} {110405} (\bibinfo {year}
  {2006})}\BibitemShut {NoStop}%
\bibitem [{\citenamefont {Yao}\ and\ \citenamefont
  {Qi}(2010)}]{PhysRevLett.105.080501}%
  \BibitemOpen
  \bibfield  {author} {\bibinfo {author} {\bibfnamefont {H.}~\bibnamefont
  {Yao}}\ and\ \bibinfo {author} {\bibfnamefont {X.-L.}\ \bibnamefont {Qi}},\
  }\href {\doibase 10.1103/PhysRevLett.105.080501} {\bibfield  {journal}
  {\bibinfo  {journal} {Phys. Rev. Lett.}\ }\textbf {\bibinfo {volume} {105}},\
  \bibinfo {pages} {080501} (\bibinfo {year} {2010})}\BibitemShut {NoStop}%
\bibitem [{\citenamefont {Jiang}\ \emph {et~al.}(2012)\citenamefont {Jiang},
  \citenamefont {Wang},\ and\ \citenamefont {Balents}}]{Jiang2012}%
  \BibitemOpen
  \bibfield  {author} {\bibinfo {author} {\bibfnamefont {H.-C.}\ \bibnamefont
  {Jiang}}, \bibinfo {author} {\bibfnamefont {Z.}~\bibnamefont {Wang}}, \ and\
  \bibinfo {author} {\bibfnamefont {L.}~\bibnamefont {Balents}},\ }\href
  {\doibase 10.1038/nphys2465} {\bibfield  {journal} {\bibinfo  {journal} {Nat.
  Phys.}\ }\textbf {\bibinfo {volume} {8}},\ \bibinfo {pages} {902} (\bibinfo
  {year} {2012})}\BibitemShut {NoStop}%
\bibitem [{\citenamefont {Miao}\ and\ \citenamefont
  {Barthel}(2022)}]{Miao2022eigenstate}%
  \BibitemOpen
  \bibfield  {author} {\bibinfo {author} {\bibfnamefont {Q.}~\bibnamefont
  {Miao}}\ and\ \bibinfo {author} {\bibfnamefont {T.}~\bibnamefont {Barthel}},\
  }\href {\doibase 10.22331/q-2022-02-02-642} {\bibfield  {journal} {\bibinfo
  {journal} {{Quantum}}\ }\textbf {\bibinfo {volume} {6}},\ \bibinfo {pages}
  {642} (\bibinfo {year} {2022})}\BibitemShut {NoStop}%
\bibitem [{\citenamefont {van Enk}\ and\ \citenamefont
  {Beenakker}(2012)}]{PhysRevLett.108.110503}%
  \BibitemOpen
  \bibfield  {author} {\bibinfo {author} {\bibfnamefont {S.~J.}\ \bibnamefont
  {van Enk}}\ and\ \bibinfo {author} {\bibfnamefont {C.~W.~J.}\ \bibnamefont
  {Beenakker}},\ }\href {\doibase 10.1103/PhysRevLett.108.110503} {\bibfield
  {journal} {\bibinfo  {journal} {Phys. Rev. Lett.}\ }\textbf {\bibinfo
  {volume} {108}},\ \bibinfo {pages} {110503} (\bibinfo {year}
  {2012})}\BibitemShut {NoStop}%
\bibitem [{\citenamefont {Elben}\ \emph {et~al.}(2018)\citenamefont {Elben},
  \citenamefont {Vermersch}, \citenamefont {Dalmonte}, \citenamefont {Cirac},\
  and\ \citenamefont {Zoller}}]{PhysRevLett.120.050406}%
  \BibitemOpen
  \bibfield  {author} {\bibinfo {author} {\bibfnamefont {A.}~\bibnamefont
  {Elben}}, \bibinfo {author} {\bibfnamefont {B.}~\bibnamefont {Vermersch}},
  \bibinfo {author} {\bibfnamefont {M.}~\bibnamefont {Dalmonte}}, \bibinfo
  {author} {\bibfnamefont {J.~I.}\ \bibnamefont {Cirac}}, \ and\ \bibinfo
  {author} {\bibfnamefont {P.}~\bibnamefont {Zoller}},\ }\href {\doibase
  10.1103/PhysRevLett.120.050406} {\bibfield  {journal} {\bibinfo  {journal}
  {Phys. Rev. Lett.}\ }\textbf {\bibinfo {volume} {120}},\ \bibinfo {pages}
  {050406} (\bibinfo {year} {2018})}\BibitemShut {NoStop}%
\bibitem [{\citenamefont {Satzinger}\ \emph {et~al.}(2021)\citenamefont
  {Satzinger}, \citenamefont {Liu}, \citenamefont {Smith}, \citenamefont
  {Knapp}, \citenamefont {Newman}, \citenamefont {Jones}, \citenamefont {Chen},
  \citenamefont {Quintana}, \citenamefont {Mi},\ and\ \citenamefont {\textit{et
  al.}}}]{doi:10.1126/science.abi8378}%
  \BibitemOpen
  \bibfield  {author} {\bibinfo {author} {\bibfnamefont {K.~J.}\ \bibnamefont
  {Satzinger}}, \bibinfo {author} {\bibfnamefont {Y.-J.}\ \bibnamefont {Liu}},
  \bibinfo {author} {\bibfnamefont {A.}~\bibnamefont {Smith}}, \bibinfo
  {author} {\bibfnamefont {C.}~\bibnamefont {Knapp}}, \bibinfo {author}
  {\bibfnamefont {M.}~\bibnamefont {Newman}}, \bibinfo {author} {\bibfnamefont
  {C.}~\bibnamefont {Jones}}, \bibinfo {author} {\bibfnamefont
  {Z.}~\bibnamefont {Chen}}, \bibinfo {author} {\bibfnamefont {C.}~\bibnamefont
  {Quintana}}, \bibinfo {author} {\bibfnamefont {X.}~\bibnamefont {Mi}}, \ and\
  \bibinfo {author} {\bibfnamefont {A.~D.}\ \bibnamefont {\textit{et al.}}},\
  }\href {\doibase 10.1126/science.abi8378} {\bibfield  {journal} {\bibinfo
  {journal} {Science}\ }\textbf {\bibinfo {volume} {374}},\ \bibinfo {pages}
  {1237} (\bibinfo {year} {2021})}\BibitemShut {NoStop}%
\bibitem [{\citenamefont {Cohen}\ \emph {et~al.}(2016)\citenamefont {Cohen},
  \citenamefont {Yukalov},\ and\ \citenamefont {Ziegler}}]{cohen16}%
   \BibitemOpen
  \bibfield  {author} {\bibinfo {author} {\bibfnamefont {D.}~\bibnamefont
  {Cohen}}, \bibinfo {author} {\bibfnamefont {V.~I.}\ \bibnamefont {Yukalov}},
  \ and\ \bibinfo {author} {\bibfnamefont {K.}~\bibnamefont {Ziegler}},\ }\href
  {\doibase 10.1103/PhysRevA.93.042101} {\bibfield  {journal} {\bibinfo
  {journal} {Phys. Rev. A}\ }\textbf {\bibinfo {volume} {93}},\ \bibinfo
  {pages} {042101} (\bibinfo {year} {2016})}\BibitemShut {NoStop}%
%%
\bibitem [{\citenamefont {Ziegler}(2021)}]{symmetry21}%
   \BibitemOpen
  \bibfield  {author} {\bibinfo {author} {\bibfnamefont {K.}~\bibnamefont
  {Ziegler}},\ }\href {\doibase 10.3390/sym13101796} {\bibfield  
  {journal} {\bibinfo {journal} {Symmetry}\ }\textbf {\bibinfo {volume} {13}},
  \ \bibinfo  {pages} {1796} (\bibinfo {year} {2021})}\BibitemShut {NoStop}%
%%
\bibitem [{\citenamefont {Anderson}(1958)}]{PhysRev.109.1492}%
  \BibitemOpen
  \bibfield  {author} {\bibinfo {author} {\bibfnamefont {P.~W.}\ \bibnamefont
  {Anderson}},\ }\href {\doibase 10.1103/PhysRev.109.1492} {\bibfield
  {journal} {\bibinfo  {journal} {Phys. Rev.}\ }\textbf {\bibinfo {volume}
  {109}},\ \bibinfo {pages} {1492} (\bibinfo {year} {1958})}\BibitemShut
  {NoStop}%
\bibitem [{\citenamefont {Huse}\ \emph {et~al.}(2013)\citenamefont {Huse},
  \citenamefont {Nandkishore}, \citenamefont {Oganesyan}, \citenamefont {Pal},\
  and\ \citenamefont {Sondhi}}]{PhysRevB.88.014206}%
  \BibitemOpen
  \bibfield  {author} {\bibinfo {author} {\bibfnamefont {D.~A.}\ \bibnamefont
  {Huse}}, \bibinfo {author} {\bibfnamefont {R.}~\bibnamefont {Nandkishore}},
  \bibinfo {author} {\bibfnamefont {V.}~\bibnamefont {Oganesyan}}, \bibinfo
  {author} {\bibfnamefont {A.}~\bibnamefont {Pal}}, \ and\ \bibinfo {author}
  {\bibfnamefont {S.~L.}\ \bibnamefont {Sondhi}},\ }\href {\doibase
  10.1103/PhysRevB.88.014206} {\bibfield  {journal} {\bibinfo  {journal} {Phys.
  Rev. B}\ }\textbf {\bibinfo {volume} {88}},\ \bibinfo {pages} {014206}
  (\bibinfo {year} {2013})}\BibitemShut {NoStop}%
\bibitem [{\citenamefont {Milburn}\ \emph {et~al.}(1997)\citenamefont
  {Milburn}, \citenamefont {Corney}, \citenamefont {Wright},\ and\
  \citenamefont {Walls}}]{PhysRevA.55.4318}%
  \BibitemOpen
  \bibfield  {author} {\bibinfo {author} {\bibfnamefont {G.~J.}\ \bibnamefont
  {Milburn}}, \bibinfo {author} {\bibfnamefont {J.}~\bibnamefont {Corney}},
  \bibinfo {author} {\bibfnamefont {E.~M.}\ \bibnamefont {Wright}}, \ and\
  \bibinfo {author} {\bibfnamefont {D.~F.}\ \bibnamefont {Walls}},\ }\href
  {\doibase 10.1103/PhysRevA.55.4318} {\bibfield  {journal} {\bibinfo
  {journal} {Phys. Rev. A}\ }\textbf {\bibinfo {volume} {55}},\ \bibinfo
  {pages} {4318} (\bibinfo {year} {1997})}\BibitemShut {NoStop}%
\bibitem [{\citenamefont {Chuchem}\ \emph {et~al.}(2010)\citenamefont
  {Chuchem}, \citenamefont {Smith-Mannschott}, \citenamefont {Hiller},
  \citenamefont {Kottos}, \citenamefont {Vardi},\ and\ \citenamefont
  {Cohen}}]{PhysRevA.82.053617}%
  \BibitemOpen
  \bibfield  {author} {\bibinfo {author} {\bibfnamefont {M.}~\bibnamefont
  {Chuchem}}, \bibinfo {author} {\bibfnamefont {K.}~\bibnamefont
  {Smith-Mannschott}}, \bibinfo {author} {\bibfnamefont {M.}~\bibnamefont
  {Hiller}}, \bibinfo {author} {\bibfnamefont {T.}~\bibnamefont {Kottos}},
  \bibinfo {author} {\bibfnamefont {A.}~\bibnamefont {Vardi}}, \ and\ \bibinfo
  {author} {\bibfnamefont {D.}~\bibnamefont {Cohen}},\ }\href {\doibase
  10.1103/PhysRevA.82.053617} {\bibfield  {journal} {\bibinfo  {journal} {Phys.
  Rev. A}\ }\textbf {\bibinfo {volume} {82}},\ \bibinfo {pages} {053617}
  (\bibinfo {year} {2010})}\BibitemShut {NoStop}%
\bibitem [{\citenamefont {Sinha}\ and\ \citenamefont
  {Sinha}(2020)}]{PhysRevLett.125.134101}%
  \BibitemOpen
  \bibfield  {author} {\bibinfo {author} {\bibfnamefont {S.}~\bibnamefont
  {Sinha}}\ and\ \bibinfo {author} {\bibfnamefont {S.}~\bibnamefont {Sinha}},\
  }\href {\doibase 10.1103/PhysRevLett.125.134101} {\bibfield  {journal}
  {\bibinfo  {journal} {Phys. Rev. Lett.}\ }\textbf {\bibinfo {volume} {125}},\
  \bibinfo {pages} {134101} (\bibinfo {year} {2020})}\BibitemShut {NoStop}%
\bibitem [{\citenamefont {Wimberger}\ \emph {et~al.}(2021)\citenamefont
  {Wimberger}, \citenamefont {Manganelli}, \citenamefont {Brollo},\ and\
  \citenamefont {Salasnich}}]{PhysRevA.103.023326}%
  \BibitemOpen
  \bibfield  {author} {\bibinfo {author} {\bibfnamefont {S.}~\bibnamefont
  {Wimberger}}, \bibinfo {author} {\bibfnamefont {G.}~\bibnamefont
  {Manganelli}}, \bibinfo {author} {\bibfnamefont {A.}~\bibnamefont {Brollo}},
  \ and\ \bibinfo {author} {\bibfnamefont {L.}~\bibnamefont {Salasnich}},\
  }\href {\doibase 10.1103/PhysRevA.103.023326} {\bibfield  {journal} {\bibinfo
   {journal} {Phys. Rev. A}\ }\textbf {\bibinfo {volume} {103}},\ \bibinfo
  {pages} {023326} (\bibinfo {year} {2021})}\BibitemShut {NoStop}%
\bibitem [{\citenamefont {Bar-Gill}\ \emph {et~al.}(2009)\citenamefont
  {Bar-Gill}, \citenamefont {Kurizki}, \citenamefont {Oberthaler},\ and\
  \citenamefont {Davidson}}]{PhysRevA.80.053613}%
  \BibitemOpen
  \bibfield  {author} {\bibinfo {author} {\bibfnamefont {N.}~\bibnamefont
  {Bar-Gill}}, \bibinfo {author} {\bibfnamefont {G.}~\bibnamefont {Kurizki}},
  \bibinfo {author} {\bibfnamefont {M.}~\bibnamefont {Oberthaler}}, \ and\
  \bibinfo {author} {\bibfnamefont {N.}~\bibnamefont {Davidson}},\ }\href
  {\doibase 10.1103/PhysRevA.80.053613} {\bibfield  {journal} {\bibinfo
  {journal} {Phys. Rev. A}\ }\textbf {\bibinfo {volume} {80}},\ \bibinfo
  {pages} {053613} (\bibinfo {year} {2009})}\BibitemShut {NoStop}%
\bibitem [{\citenamefont {Abbarchi}\ \emph {et~al.}(2013)\citenamefont
  {Abbarchi}, \citenamefont {Amo}, \citenamefont {Sala}, \citenamefont
  {Solnyshkov}, \citenamefont {Flayac}, \citenamefont {Ferrier}, \citenamefont
  {Sagnes}, \citenamefont {Galopin}, \citenamefont {Lema{\^i}tre},
  \citenamefont {Malpuech},\ and\ \citenamefont {Bloch}}]{Abbarchi2013}%
  \BibitemOpen
  \bibfield  {author} {\bibinfo {author} {\bibfnamefont {M.}~\bibnamefont
  {Abbarchi}}, \bibinfo {author} {\bibfnamefont {A.}~\bibnamefont {Amo}},
  \bibinfo {author} {\bibfnamefont {V.~G.}\ \bibnamefont {Sala}}, \bibinfo
  {author} {\bibfnamefont {D.~D.}\ \bibnamefont {Solnyshkov}}, \bibinfo
  {author} {\bibfnamefont {H.}~\bibnamefont {Flayac}}, \bibinfo {author}
  {\bibfnamefont {L.}~\bibnamefont {Ferrier}}, \bibinfo {author} {\bibfnamefont
  {I.}~\bibnamefont {Sagnes}}, \bibinfo {author} {\bibfnamefont
  {E.}~\bibnamefont {Galopin}}, \bibinfo {author} {\bibfnamefont
  {A.}~\bibnamefont {Lema{\^i}tre}}, \bibinfo {author} {\bibfnamefont
  {G.}~\bibnamefont {Malpuech}}, \ and\ \bibinfo {author} {\bibfnamefont
  {J.}~\bibnamefont {Bloch}},\ }\href {\doibase 10.1038/nphys2609} {\bibfield
  {journal} {\bibinfo  {journal} {Nat. Phys.}\ }\textbf {\bibinfo {volume}
  {9}},\ \bibinfo {pages} {275} (\bibinfo {year} {2013})}\BibitemShut {NoStop}%
\bibitem [{\citenamefont {Spagnolli}\ \emph {et~al.}(2017)\citenamefont
  {Spagnolli}, \citenamefont {Semeghini}, \citenamefont {Masi}, \citenamefont
  {Ferioli}, \citenamefont {Trenkwalder}, \citenamefont {Coop}, \citenamefont
  {Landini}, \citenamefont {Pezz\`e}, \citenamefont {Modugno}, \citenamefont
  {Inguscio}, \citenamefont {Smerzi},\ and\ \citenamefont
  {Fattori}}]{PhysRevLett.118.230403}%
  \BibitemOpen
  \bibfield  {author} {\bibinfo {author} {\bibfnamefont {G.}~\bibnamefont
  {Spagnolli}}, \bibinfo {author} {\bibfnamefont {G.}~\bibnamefont
  {Semeghini}}, \bibinfo {author} {\bibfnamefont {L.}~\bibnamefont {Masi}},
  \bibinfo {author} {\bibfnamefont {G.}~\bibnamefont {Ferioli}}, \bibinfo
  {author} {\bibfnamefont {A.}~\bibnamefont {Trenkwalder}}, \bibinfo {author}
  {\bibfnamefont {S.}~\bibnamefont {Coop}}, \bibinfo {author} {\bibfnamefont
  {M.}~\bibnamefont {Landini}}, \bibinfo {author} {\bibfnamefont
  {L.}~\bibnamefont {Pezz\`e}}, \bibinfo {author} {\bibfnamefont
  {G.}~\bibnamefont {Modugno}}, \bibinfo {author} {\bibfnamefont
  {M.}~\bibnamefont {Inguscio}}, \bibinfo {author} {\bibfnamefont
  {A.}~\bibnamefont {Smerzi}}, \ and\ \bibinfo {author} {\bibfnamefont
  {M.}~\bibnamefont {Fattori}},\ }\href {\doibase
  10.1103/PhysRevLett.118.230403} {\bibfield  {journal} {\bibinfo  {journal}
  {Phys. Rev. Lett.}\ }\textbf {\bibinfo {volume} {118}},\ \bibinfo {pages}
  {230403} (\bibinfo {year} {2017})}\BibitemShut {NoStop}%
\bibitem [{\citenamefont {Pigneur}\ \emph {et~al.}(2018)\citenamefont
  {Pigneur}, \citenamefont {Berrada}, \citenamefont {Bonneau}, \citenamefont
  {Schumm}, \citenamefont {Demler},\ and\ \citenamefont
  {Schmiedmayer}}]{PhysRevLett.120.173601}%
  \BibitemOpen
  \bibfield  {author} {\bibinfo {author} {\bibfnamefont {M.}~\bibnamefont
  {Pigneur}}, \bibinfo {author} {\bibfnamefont {T.}~\bibnamefont {Berrada}},
  \bibinfo {author} {\bibfnamefont {M.}~\bibnamefont {Bonneau}}, \bibinfo
  {author} {\bibfnamefont {T.}~\bibnamefont {Schumm}}, \bibinfo {author}
  {\bibfnamefont {E.}~\bibnamefont {Demler}}, \ and\ \bibinfo {author}
  {\bibfnamefont {J.}~\bibnamefont {Schmiedmayer}},\ }\href {\doibase
  10.1103/PhysRevLett.120.173601} {\bibfield  {journal} {\bibinfo  {journal}
  {Phys. Rev. Lett.}\ }\textbf {\bibinfo {volume} {120}},\ \bibinfo {pages}
  {173601} (\bibinfo {year} {2018})}\BibitemShut {NoStop}%
\bibitem [{\citenamefont {Mennemann}\ \emph {et~al.}(2021)\citenamefont
  {Mennemann}, \citenamefont {Mazets}, \citenamefont {Pigneur}, \citenamefont
  {Stimming}, \citenamefont {Mauser}, \citenamefont {Schmiedmayer},\ and\
  \citenamefont {Erne}}]{PhysRevResearch.3.023197}%
  \BibitemOpen
  \bibfield  {author} {\bibinfo {author} {\bibfnamefont {J.-F.}\ \bibnamefont
  {Mennemann}}, \bibinfo {author} {\bibfnamefont {I.~E.}\ \bibnamefont
  {Mazets}}, \bibinfo {author} {\bibfnamefont {M.}~\bibnamefont {Pigneur}},
  \bibinfo {author} {\bibfnamefont {H.~P.}\ \bibnamefont {Stimming}}, \bibinfo
  {author} {\bibfnamefont {N.~J.}\ \bibnamefont {Mauser}}, \bibinfo {author}
  {\bibfnamefont {J.}~\bibnamefont {Schmiedmayer}}, \ and\ \bibinfo {author}
  {\bibfnamefont {S.}~\bibnamefont {Erne}},\ }\href {\doibase
  10.1103/PhysRevResearch.3.023197} {\bibfield  {journal} {\bibinfo  {journal}
  {Phys. Rev. Research}\ }\textbf {\bibinfo {volume} {3}},\ \bibinfo {pages}
  {023197} (\bibinfo {year} {2021})}\BibitemShut {NoStop}%
\bibitem [{\citenamefont {Kim}\ and\ \citenamefont
  {Huse}(2013)}]{PhysRevLett.111.127205}%
  \BibitemOpen
  \bibfield  {author} {\bibinfo {author} {\bibfnamefont {H.}~\bibnamefont
  {Kim}}\ and\ \bibinfo {author} {\bibfnamefont {D.~A.}\ \bibnamefont {Huse}},\
  }\href {\doibase 10.1103/PhysRevLett.111.127205} {\bibfield  {journal}
  {\bibinfo  {journal} {Phys. Rev. Lett.}\ }\textbf {\bibinfo {volume} {111}},\
  \bibinfo {pages} {127205} (\bibinfo {year} {2013})}\BibitemShut {NoStop}%
\bibitem [{\citenamefont {Wigner}(1955)}]{10.2307/1970079}%
  \BibitemOpen
  \bibfield  {author} {\bibinfo {author} {\bibfnamefont {E.~P.}\ \bibnamefont
  {Wigner}},\ }\href {http://www.jstor.org/stable/1970079} {\bibfield
  {journal} {\bibinfo  {journal} {Ann. Math.}\ }\textbf {\bibinfo {volume}
  {62}},\ \bibinfo {pages} {548} (\bibinfo {year} {1955})}\BibitemShut
  {NoStop}%
\bibitem [{\citenamefont {Porter}\ and\ \citenamefont
  {Thomas}(1956)}]{PhysRev.104.483}%
  \BibitemOpen
  \bibfield  {author} {\bibinfo {author} {\bibfnamefont {C.~E.}\ \bibnamefont
  {Porter}}\ and\ \bibinfo {author} {\bibfnamefont {R.~G.}\ \bibnamefont
  {Thomas}},\ }\href {\doibase 10.1103/PhysRev.104.483} {\bibfield  {journal}
  {\bibinfo  {journal} {Phys. Rev.}\ }\textbf {\bibinfo {volume} {104}},\
  \bibinfo {pages} {483} (\bibinfo {year} {1956})}\BibitemShut {NoStop}%
\bibitem [{\citenamefont {Dyson}(1962)}]{doi:10.1063/1.1703773}%
  \BibitemOpen
  \bibfield  {author} {\bibinfo {author} {\bibfnamefont {F.~J.}\ \bibnamefont
  {Dyson}},\ }\href {\doibase 10.1063/1.1703773} {\bibfield  {journal}
  {\bibinfo  {journal} {J. Math. Phys.}\ }\textbf {\bibinfo {volume} {3}},\
  \bibinfo {pages} {140} (\bibinfo {year} {1962})}\BibitemShut {NoStop}%
\bibitem [{\citenamefont {Wigner}(1967)}]{10.2307/2027409}%
  \BibitemOpen
  \bibfield  {author} {\bibinfo {author} {\bibfnamefont {E.~P.}\ \bibnamefont
  {Wigner}},\ }\href {http://www.jstor.org/stable/2027409} {\bibfield
  {journal} {\bibinfo  {journal} {SIAM Rev.}\ }\textbf {\bibinfo {volume}
  {9}},\ \bibinfo {pages} {1} (\bibinfo {year} {1967})}\BibitemShut {NoStop}%
\bibitem [{\citenamefont {Mehta}(2004)}]{mehta2004random}%
  \BibitemOpen
  \bibfield  {author} {\bibinfo {author} {\bibfnamefont {M.~L.}\ \bibnamefont
  {Mehta}},\ }\href@noop {} {\emph {\bibinfo {title} {Random Matrices}}}\
  (\bibinfo  {publisher} {Elsevier, New York},\ \bibinfo {year}
  {2004})\BibitemShut {NoStop}%
\bibitem [{\citenamefont {Beenakker}(1997)}]{RevModPhys.69.731}%
  \BibitemOpen
  \bibfield  {author} {\bibinfo {author} {\bibfnamefont {C.~W.~J.}\
  \bibnamefont {Beenakker}},\ }\href {\doibase 10.1103/RevModPhys.69.731}
  {\bibfield  {journal} {\bibinfo  {journal} {Rev. Mod. Phys.}\ }\textbf
  {\bibinfo {volume} {69}},\ \bibinfo {pages} {731} (\bibinfo {year}
  {1997})}\BibitemShut {NoStop}%
\bibitem [{\citenamefont {Li}\ and\ \citenamefont
  {Haldane}(2008)}]{PhysRevLett.101.010504}%
  \BibitemOpen
  \bibfield  {author} {\bibinfo {author} {\bibfnamefont {H.}~\bibnamefont
  {Li}}\ and\ \bibinfo {author} {\bibfnamefont {F.~D.~M.}\ \bibnamefont
  {Haldane}},\ }\href {\doibase 10.1103/PhysRevLett.101.010504} {\bibfield
  {journal} {\bibinfo  {journal} {Phys. Rev. Lett.}\ }\textbf {\bibinfo
  {volume} {101}},\ \bibinfo {pages} {010504} (\bibinfo {year}
  {2008})}\BibitemShut {NoStop}%
\bibitem [{\citenamefont {Gong}\ and\ \citenamefont
  {Ueda}(2018)}]{PhysRevLett.121.250601}%
  \BibitemOpen
  \bibfield  {author} {\bibinfo {author} {\bibfnamefont {Z.}~\bibnamefont
  {Gong}}\ and\ \bibinfo {author} {\bibfnamefont {M.}~\bibnamefont {Ueda}},\
  }\href {\doibase 10.1103/PhysRevLett.121.250601} {\bibfield  {journal}
  {\bibinfo  {journal} {Phys. Rev. Lett.}\ }\textbf {\bibinfo {volume} {121}},\
  \bibinfo {pages} {250601} (\bibinfo {year} {2018})}\BibitemShut {NoStop}%
\bibitem [{\citenamefont {Parisen~Toldin}\ and\ \citenamefont
  {Assaad}(2018)}]{PhysRevLett.121.200602}%
  \BibitemOpen
  \bibfield  {author} {\bibinfo {author} {\bibfnamefont {F.}~\bibnamefont
  {Parisen~Toldin}}\ and\ \bibinfo {author} {\bibfnamefont {F.~F.}\
  \bibnamefont {Assaad}},\ }\href {\doibase 10.1103/PhysRevLett.121.200602}
  {\bibfield  {journal} {\bibinfo  {journal} {Phys. Rev. Lett.}\ }\textbf
  {\bibinfo {volume} {121}},\ \bibinfo {pages} {200602} (\bibinfo {year}
  {2018})}\BibitemShut {NoStop}%
\bibitem [{\citenamefont {Fries}\ and\ \citenamefont
  {Reyes}(2019)}]{PhysRevLett.123.211603}%
  \BibitemOpen
  \bibfield  {author} {\bibinfo {author} {\bibfnamefont {P.}~\bibnamefont
  {Fries}}\ and\ \bibinfo {author} {\bibfnamefont {I.~A.}\ \bibnamefont
  {Reyes}},\ }\href {\doibase 10.1103/PhysRevLett.123.211603} {\bibfield
  {journal} {\bibinfo  {journal} {Phys. Rev. Lett.}\ }\textbf {\bibinfo
  {volume} {123}},\ \bibinfo {pages} {211603} (\bibinfo {year}
  {2019})}\BibitemShut {NoStop}%
\bibitem [{\citenamefont {Riera-Campeny}\ \emph {et~al.}(2021)\citenamefont
  {Riera-Campeny}, \citenamefont {Sanpera},\ and\ \citenamefont
  {Strasberg}}]{PRXQuantum.2.010340}%
  \BibitemOpen
  \bibfield  {author} {\bibinfo {author} {\bibfnamefont {A.}~\bibnamefont
  {Riera-Campeny}}, \bibinfo {author} {\bibfnamefont {A.}~\bibnamefont
  {Sanpera}}, \ and\ \bibinfo {author} {\bibfnamefont {P.}~\bibnamefont
  {Strasberg}},\ }\href {\doibase 10.1103/PRXQuantum.2.010340} {\bibfield
  {journal} {\bibinfo  {journal} {PRX Quantum}\ }\textbf {\bibinfo {volume}
  {2}},\ \bibinfo {pages} {010340} (\bibinfo {year} {2021})}\BibitemShut
  {NoStop}%
  \bibitem{Ziegler2012}
 K. Ziegler, Laser Physics {\bf 22}, 331 (2012)

\bibitem{Ziegler2017}
K. Ziegler, International Journal of Modern Physics B {\bf 31}, 1750255 (2017)
\end{thebibliography}
%\bibliographystyle{apsrev4-2.bst}

\iffalse

\fi


\end{document}